\documentclass[a4paper]{article}
\usepackage[OT1]{fontenc}
\usepackage[english]{babel}
\usepackage[utf8]{inputenc}
\usepackage{amsmath,amssymb,amsfonts,mathrsfs}
\usepackage{amsthm}
\usepackage{graphicx}
\usepackage{caption}
\usepackage{subcaption}
\usepackage{multirow}
\usepackage{enumitem}
\usepackage{varioref}
\usepackage{datetime}
\usepackage{mathtools}
\usepackage{array}
\usepackage{xcolor}

\title{Deep Investing in Kyle's Single Period Model}
\author{Paul Friedrich and Josef Teichmann}
\date{\today}

\newcommand{\E}{\mathbb{E}}

\renewcommand{\epsilon}{\ensuremath\varepsilon}

\renewcommand{\phi}{\ensuremath{\varphi}}

\newtheorem{thm}{Theorem}[section]

\newtheorem{defn}[thm]{Definition}

\usepackage{multicol}
\begin{document}
\maketitle

\begin{abstract}
The Kyle model describes how an equilibrium of order sizes and security prices naturally arises between a trader with insider information and the price providing market maker as they interact through a series of auctions. Ever since being introduced by Albert S. Kyle in 1985, the model has become important in the study of market microstructure models with asymmetric information. As it is well understood, it serves as an excellent opportunity to study how modern deep learning technology can be used to replicate and better understand equilibria that occur in certain market learning problems.

We model the agents in Kyle's single period setting using deep neural networks. The networks are trained by interacting following the rules and objectives as defined by Kyle. We show how the right network architectures and training methods lead to the agents' behaviour converging to the theoretical equilibrium that is predicted by Kyle's model. 
\end{abstract}

\section{Introduction}
The Kyle model, introduced by Albert S. Kyle in his 1985 paper \emph{Continuous Auctions and Insider Trading} \cite{originalkyle}, is a well known and often cited market microstructure model. It aims to describe and understand aspects of insider trading, liquidity and the value of private information to an insider. Our work has two parts. We use neural networks to replicate the single period Kyle model under Gaussian assumptions and then alter it, using the trained networks to find Kyle equilibria where tractability fails to hold. For the theoretical discussions, we will mainly be following Kyle's original paper \cite{originalkyle}. 

In the model, a risky asset is traded over one period of time in a series of auctions. Since a risk-free asset is used as a numeraire, the risk-free rate is set to zero. The market contains three actors: a risk neutral inside trader, a random noise trader and a risk neutral market maker. The inside trader possesses private and exclusive knowledge of the liquidation price of the asset, i.e., the value at which it can be sold at the end of the auction. Based on this knowledge, the insider submits a market order. The noise trader also submits a market order, but its size is random. Both orders are aggregated and sent to the market maker, who is unable to distinguish which proportion of the total order originates from each trader. This way, the noise trader provides camouflage for the inside trader, enabling the insider to make a profit at the noise trader's expense. The market maker then determines a price and clears the market. Kyle proved in \cite{originalkyle} that in the situation of a single auction or a series of independent auctions under Gaussian assumptions, an equilibrium arises between the inside trader and market maker. Market maker prices and inside trader order quantities are linear functions of their respective inputs, and their parameters can be derived explicitly. He also showed that the multi-auction model and its equilibrium converge to a continuous model and equilibrium as the time between auctions approaches zero.

The model has several important features. A crucial one is that the insider considers their own impact on the offered market prices. They need to decide how to optimally use their private information in order to maximise profit. One can also use the Kyle model to understand liquidity conditions of a market in an auction equilibrium. A liquid market is, as written in Kyle \cite{originalkyle}, characterised as one that is nearly infinitely tight (i.e., turning around a large position over a short period of time is very expensive), not infinitely deep (i.e., one does not need extremely large order flows to impact the market price) and resilient (i.e., prices tend to return to their underlying value over time). The Kyle model allows one to study these dynamics in a quantitative way. Lastly, the model shows how private trading information impacts the markets via the inside trader, who gradually uses that information and incorporates it into market prices. The market situation with its agents interacting with each other using fixed constraints and reward functions is well suited to the application of machine learning techniques. 

This paper has a two step approach. In Chapter \ref{chap:theory}, we illustrate the theoretical setting, show the central statements about the existence of an equilibrium, and formulate the theory in a way that makes it easier to translate into machine learning implementations. By replacing the agents with neural networks who then trade with each other, we can better understand the nature and dynamics of the equilibrium. Once an equilibrium is found by the agents, we can alter the parameters and constraints of the problem and introduce new aspects. These can for example include transaction costs, different price distributions or irrational behaviour by one or both of the agents. In Chapter \ref{chap:results}, we apply this approach to the version of the model where only a single auction takes place and illustrate our methods and results. 

\section{Setting}\label{chap:theory}

The single period Kyle model describes a market model where a single risky asset is traded among three agents: a risk-neutral \emph{market maker}, a \emph{noise trader} and a risk-neutral \emph{insider}. At the beginning of the period, the noise trader and insider submit their market orders (positive if buy order, negative if short sell order) to the market maker, who then determines one price for both transactions. At the end of the period, the noise trader and insider realise a profit or loss, depending on the value of the asset and initial price. We first list the definitions and assumptions within the model.

The value of the traded asset at the end of the period is denoted $z$ and assumed to be a realisation of a normally distributed random variable $Z \sim \mathcal{N}(\mu_z, \sigma_z^2)$.  The noise trader submits a market order of size $y$ to the market maker. The order is modelled as a normal random variable $Y \sim \mathcal{N}(0, \sigma_y^2)$ that is independent of $Z$ and any other elements of the model. It is assumed that the insider knows with certainty the exact value $z$ of the asset at the end of the period. They do not know the order size $y$ of the noise trader. Based on this information and with the goal to maximise his profits, the insider submits a market order of size $x=X(z)$ to the market maker. The market maker receives the combined total of market orders from the insider and noise trader, $x+y$. However, they cannot differentiate which proportion of the total order came from the insider or noise trader. The market maker uses a fixed pricing rule $P(\cdot)$ that depends only on the total market order amount $X+Y$ and the market maker's goal of exactly breaking even on their own trade. Based on the total order obtained the market price $p=P(x+y)$ is set according to the market maker's pricing rule. Afterwards they take the position $-(x+y)$ to clear the market.

Both the market maker and insider are assumed to act risk-neutral. The insider's goal is to find the value $x$ that maximises his or her expected end-of-period profit while knowing the exact end-of-period asset price $z$, i.e. 
\begin{equation}
\max_x\E[(Z-P(x+Y))x|Z=z] = \max_x (z-\E[P(x+Y)])x.\tag{2.1}\label{eq:1}
\end{equation}

For the market maker, it is assumed that the pricing rule solely depends on the total market orders. Indeed, the greater that total size is, the greater the possibility of $x$ being large, which would indicate that the insider knows that the end-of-period price $z$ is higher than the expected $\mu$. Therefore the market maker would set a higher price. The opposite also holds, the smaller the total order size is, the greater the possibility of $x$ being small, indicating that $z$ would fall below $\mu$. The market maker would then lower the offered price. Since the market maker's expected end-of-period profit is given by $-\E[(Z-P(x+y))(x+y)]$ and they have the goal to break even, the pricing rule is given by
\begin{equation}
P(x+y) = \E[Z|X + Y = x + y].\tag{2.2}\label{eq:2}
\end{equation}

The goal is now to find a situation where the insider and market maker are in an \emph{equilibrium}. 

\begin{defn}
Let $p=P(X+Y)$ be the random variable describing the market maker's price whose distribution depends solely on $X$ and $P$. Let $\pi = X\cdot(Z-p)$ be the random variable describing the inside trader's profit which depends solely on their strategy $X$ and the market maker's pricing rule $P$. An \emph{equilibrium} is given by $X$ and $P$ which satisfy the following conditions:
\begin{enumerate}[label=(\roman*)]
\item Profit Maximisation: For all trading strategies $X' \neq X$ and end-of-period prices $z$, it holds that
$$\E[\pi(X, P)|Z = z] \geq \E[\pi(X', P)|Z = z].$$
\item Market Efficiency: The random variable $p$ satisfies
$$ p(X, P) = \E[Z|X+Y].$$
\end{enumerate}
\end{defn}

Intuitively, the insider succeeds in maximising their profits given the market maker's pricing rule, the market maker succeeds in having a net profit of zero given the total market orders by noise trader and insider and both the insider and market maker expect each other's behaviour. 

Kyle argues that under the assumption that $Y$ and $Z$ are independent, an equilibrium can be found where both market maker and insider use a linear pricing and trading rule respectively. The proof of this theorem can be found in \cite{originalkyle} as the proof of Theorem 1.

\begin{thm} \label{thm:1}
There exists a unique equilibrium in which $X$ and $P$ are both linear functions. For the constants $\alpha = -\frac{\sigma_y}{\sigma_z}\mu_z,\ \beta = \frac{\sigma_y}{\sigma_z}\ \textrm{and}\ \mu = \mu_z, \lambda = \frac{1}{2}\frac{\sigma_y}{\sigma_z}$ it holds that
$$X(z) = \alpha + \beta z,\quad P(x+y) = \mu + \lambda (x+y).$$
\end{thm}

One important fact for our implementation is that for estimating the pricing rule $P(x+y) = \E[Z|X + Y = x + y]$, a maximum likelihood estimator is the 'best' course of action in the sense that it leads to maximum efficiency while being the minimum variance unbiased estimate. Since the insider uses a linear pricing rule such as $X(Z) = \alpha + \beta Z$, from the point of view of the market maker, $Z$ and $V:=Y+X=Y+\alpha+\beta Z$ are jointly normally distributed. Therefore, the maximum likelihood estimate of $\E[Z|V]$ is linear in V and in this case is actually the least squares one, i.e. the one that minimises $\E[(Z-P(V))^2]$. This provides an easy loss function for the market maker model.

\section{Methods and results}\label{chap:results}


\subsection{Network architecture, loss functions and training}

We implement Kyle's model in a deep learning framework in order to test if a pair of neural networks is able to arrive at the linear equilibrium using only the base assumptions of the model. We implement and train the models with \emph{Keras} using \emph{TensorFlow} as a backend. The models consist of a simple feed-forward layer structure where the market maker possesses two layers and the insider one layer of ten nodes each, both times connected to a single output node. As for the model defining parameters, we choose a configuration of $\mu_z = 0.5, \sigma_z = 2, \sigma_y = 1$, which we refer to as `non-centered'. We found that using large positive values of $\mu_z$ for training would lead to the insider's predictions being highly inaccurate for negative values of $z$, as the likelihood of encountering such values during training would be very low. $\mu_z = 0$ (a `centered' configuration) or close to zero produced the best results. 

All assumptions we make are in line with Kyle's model, namely:
\begin{enumerate}
\item The noise trader market order $Y$ and end-of-period price $Z$ are independent and normally distributed with the parameters we chose.
\item The insider knows the end-of-period market price $z$.
\item The market maker receives only the combined total of market orders $x+y$.
\end{enumerate}

Next, let us describe our training procedure. We run a training loop where in each loop we first train the market maker based on the most recently trained insider, then train the insider based on the market maker. The very first loop does not yet have an insider model initialized and we therefore require an initial insider order function to train the market maker. We then plot the results and begin a new training loop. In this way, we alternately train each actor with the intention of both actors gradually learning from each other until an equilibrium is reached. Before we begin training, we thus have to choose a configuration for the distributions of $Z$ and $Y$, as well as an initial insider order function. We usually choose an initial insider order function that returns values randomly, namely normally distributed like the noise trader $Y$, i.e. $X(\cdot) \sim Y$. This approach of letting the initial insider output noise uses zero outside information to train the market maker. In the following discussion, we refer to the insider and market maker networks as `I' and `M' respectively. $I(z)$ or $M(x+y)$ then denote the network outputs when given an input $z$ or $x+y$.

Each loop begins by training the market maker. We first sample from $Y$ and $Z$ independently, generating samples $(y_1,\ldots,y_N)$ and $(z_1,\ldots,z_N)$. We use the initial insider order function (first loop) or the last trained insider network (subsequent loops) to generate predictions $I(z_i):=x(z_i)$ out of the samples of Z. The $y_i$ are added to the corresponding insider orders $I(z_i)$ and result in a sample of total market orders $(I(z_1)+y_1,\ldots,I(z_N)+y_N)$. For all $i \leq N$, $I(z_i)+y_i$ is fed into the market maker neural network together with the corresponding $z_i$ (which is only used in the loss function) with batch size one. TensorFlow minimises the loss function of a whole epoch at a time. This loss is essentially the mean of each batches' individual loss. We therefore define the following loss function for the market maker:
\begin{itemize}
	\item Calculate a loss for each input $I(z_i)+y_i$:
	$$L(z_i,y_i) := (z_i-M(I(z_i)+y_i))^2.$$
	\item The total loss over one epoch of training is obtained by averaging all $L(z)$:
	$$L_\text{epoch} = \E_{(Z,Y)}[(Z-M(I(Z)+Y))^2].$$
\end{itemize}
The model is trained for a number of epochs on the sampled data. Its output, via a \verb|model.predict| method, then defines a pricing function $P(\cdot) := M(\cdot)$ for the insider to use.

Next, we train the insider. We again sample from $Y$ and $Z$ independently. The samples can be used directly for training, as the market maker appears only in the insider's loss function in what is a straightforward implementation of equation (\ref{eq:1}):
\begin{itemize}
	\item Calculate a loss for each input $z_i, y_i$:
	$$L(z_i,y_i):= -(z_i-M(I(z_i)+y_i))I(z_i).$$
	\item The total loss over one epoch of training is then given by:
	$$L_\text{epoch} = -\E_{(Z,Y)}[(Z-M(I(Z)+Y))I(Z)].$$
\end{itemize}
The insider is trained and its output defines a new order function $x(\cdot) := I(\cdot)$ that is used to train the market maker in the next iteration. In both cases we rely on neural network generalization which determines the functions outside the initial training data.

\subsection{Results}

We train the model using a centered configuration with $\mu_z = 0$. Each model is trained for three epochs within one loop, with there being a total of 20 training loops. While the models converge using as little as 2000 samples, we use 5000 samples of $z$ and $y$ for both the insider and market maker. This way, convergence happens within fewer training loops and the convergence is more stable in the sense that model predictions do not change much after first hitting an acceptable solution. 
\begin{figure}
\centering
	\begin{subfigure}[b]{0.5\textwidth}
	\raggedleft
	\includegraphics[width=0.9\linewidth]{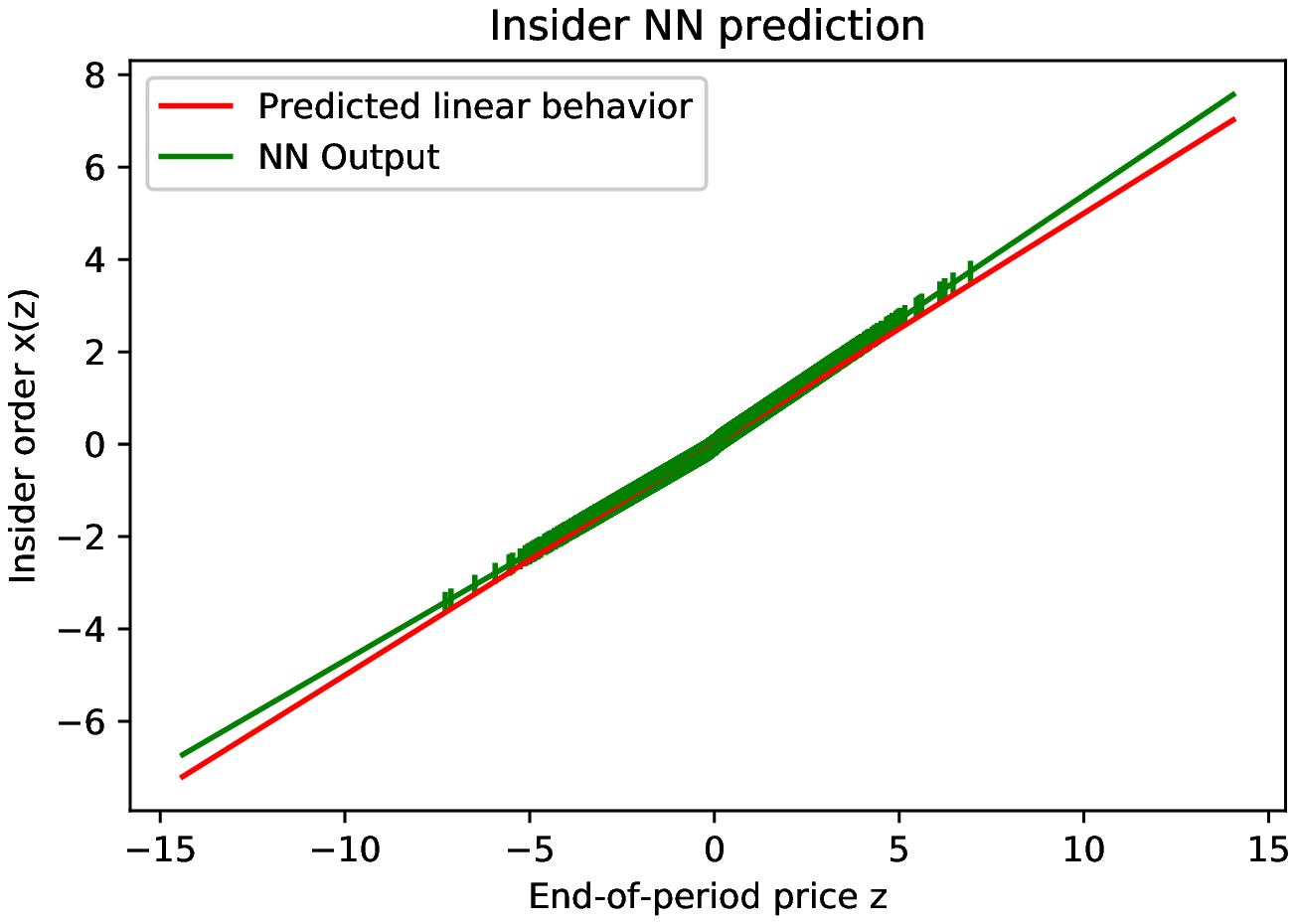}
	\label{fig:linstart_cent_I}
	\end{subfigure}%
	\begin{subfigure}[b]{0.5\textwidth}
	\raggedright 
	\includegraphics[width=0.9\linewidth]{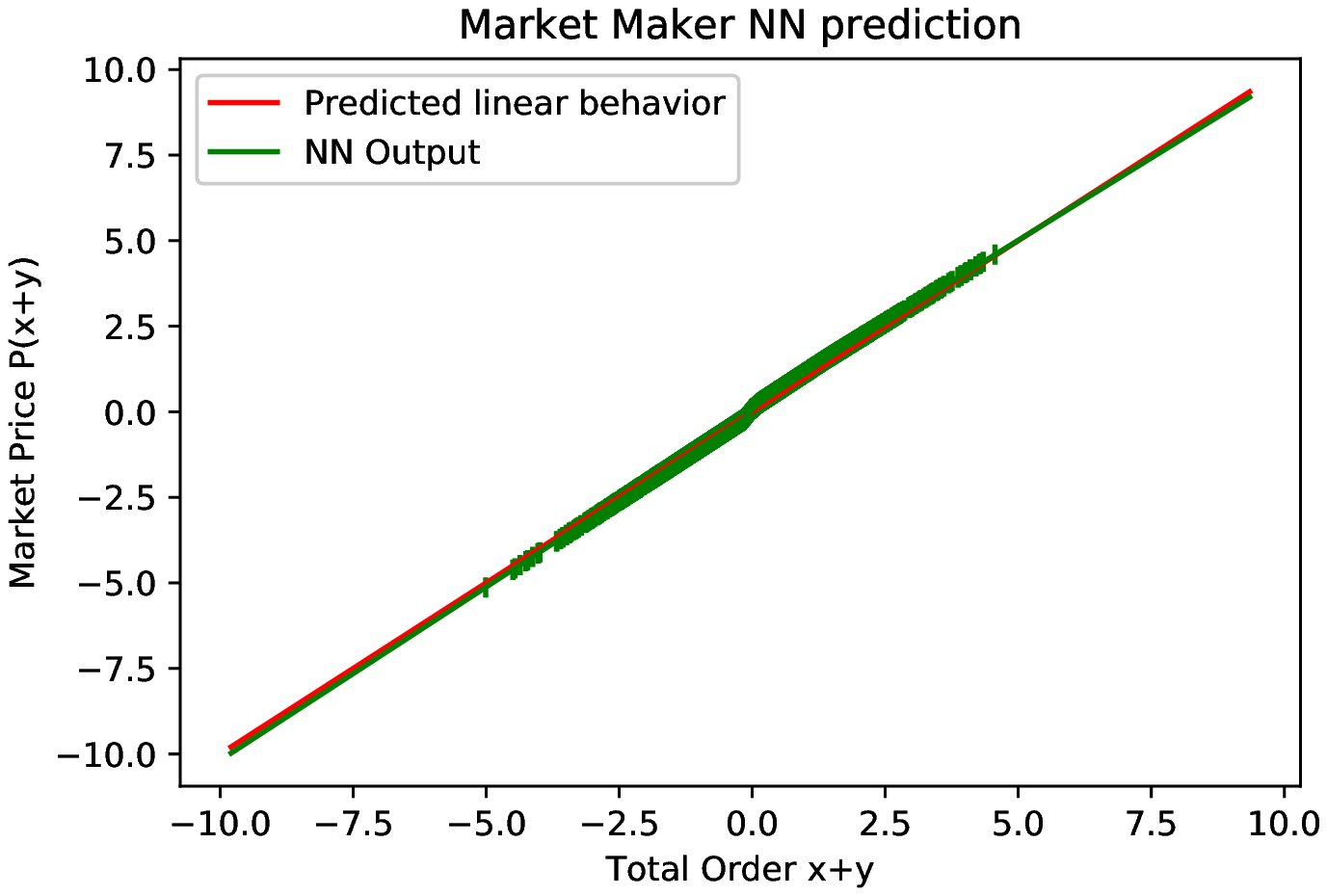}
	\label{fig:linstart_cent_MM}
	\end{subfigure}
	\begin{subfigure}[b]{0.5\textwidth}
	\raggedleft
	   \includegraphics[width=0.9\linewidth]{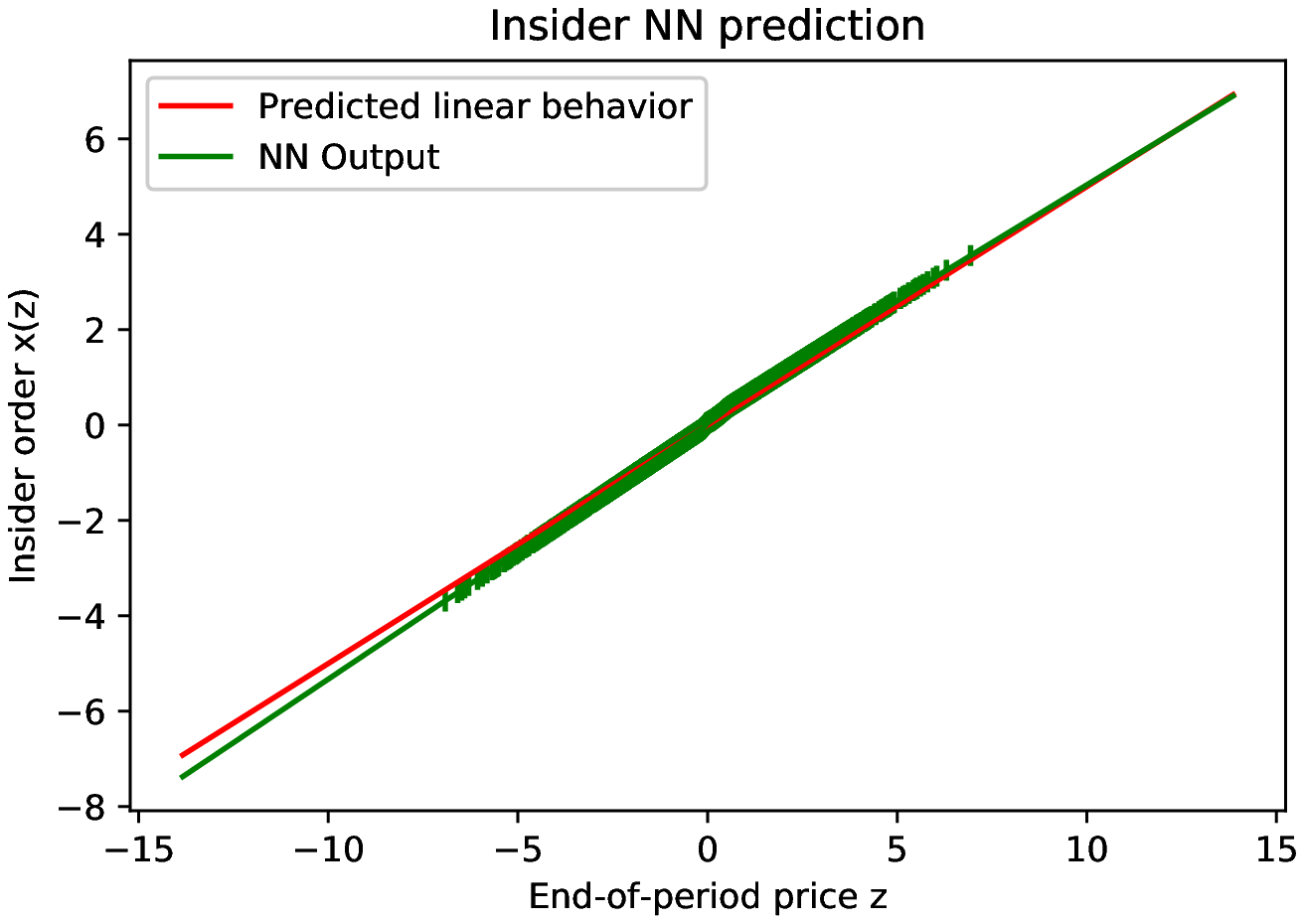}
 	   \label{fig:expstart_cent_I} 
	\end{subfigure}%
	\begin{subfigure}[b]{0.5\textwidth}
	\raggedright 
 	  \includegraphics[width=0.9\linewidth]{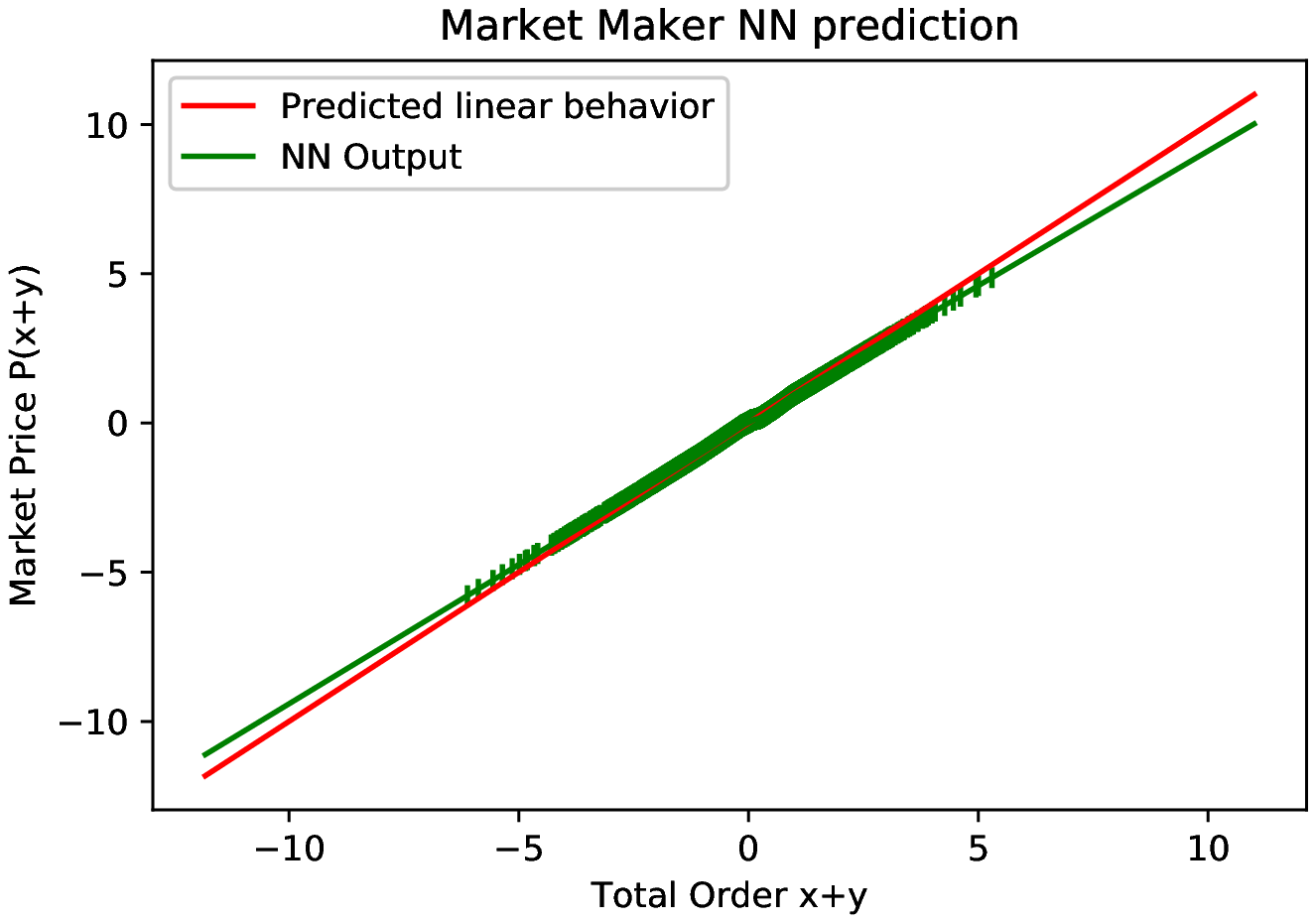}
 	  \label{fig:expstart_cent_MM}
	\end{subfigure} 
	\begin{subfigure}[b]{0.5\textwidth}
	\raggedleft
	   \includegraphics[width=0.9\linewidth]{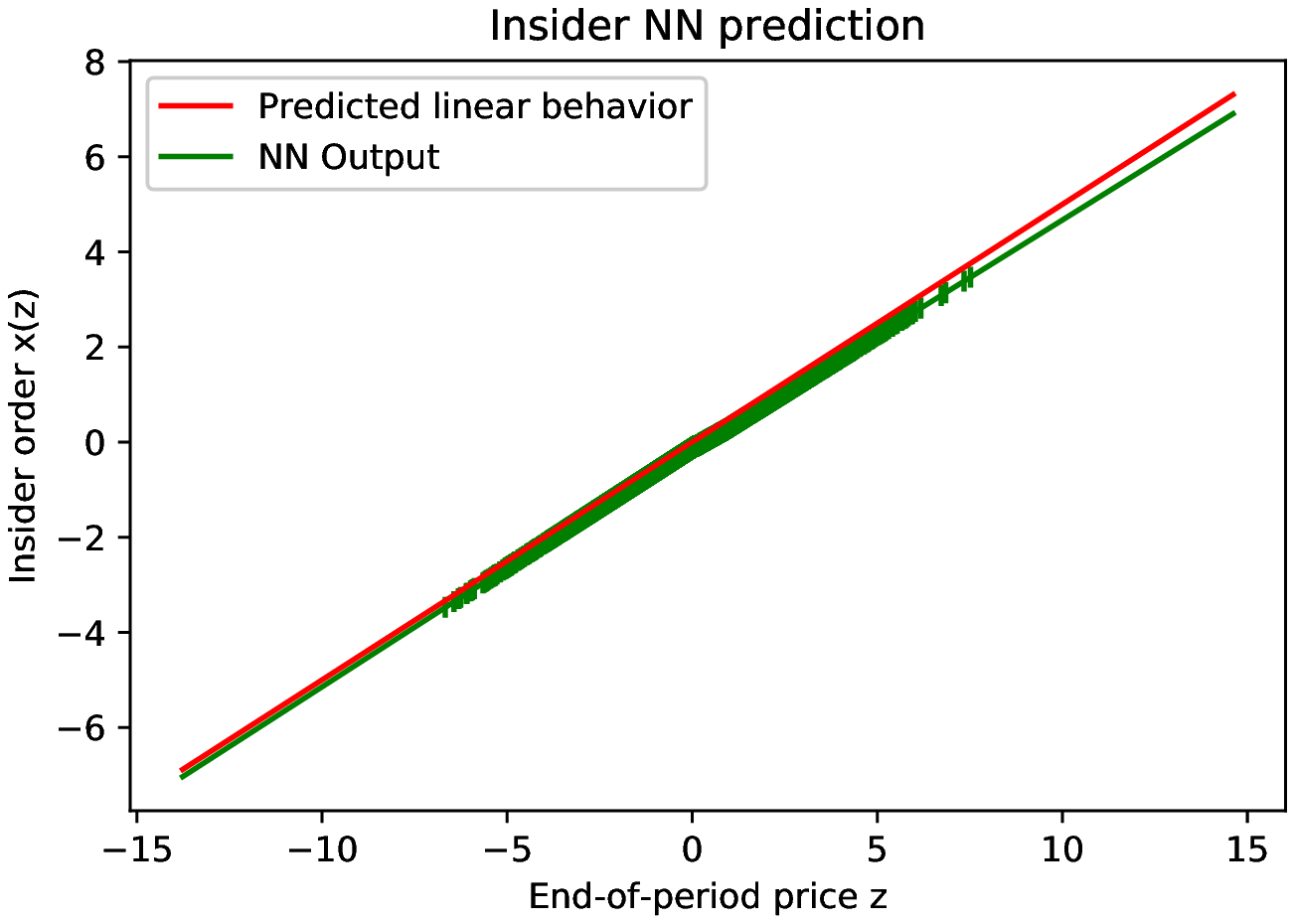}
 	   \label{fig:noisestart_cent_I} 
	\end{subfigure}%
	\begin{subfigure}[b]{0.5\textwidth}
	\raggedright 
 	  \includegraphics[width=0.9\linewidth]{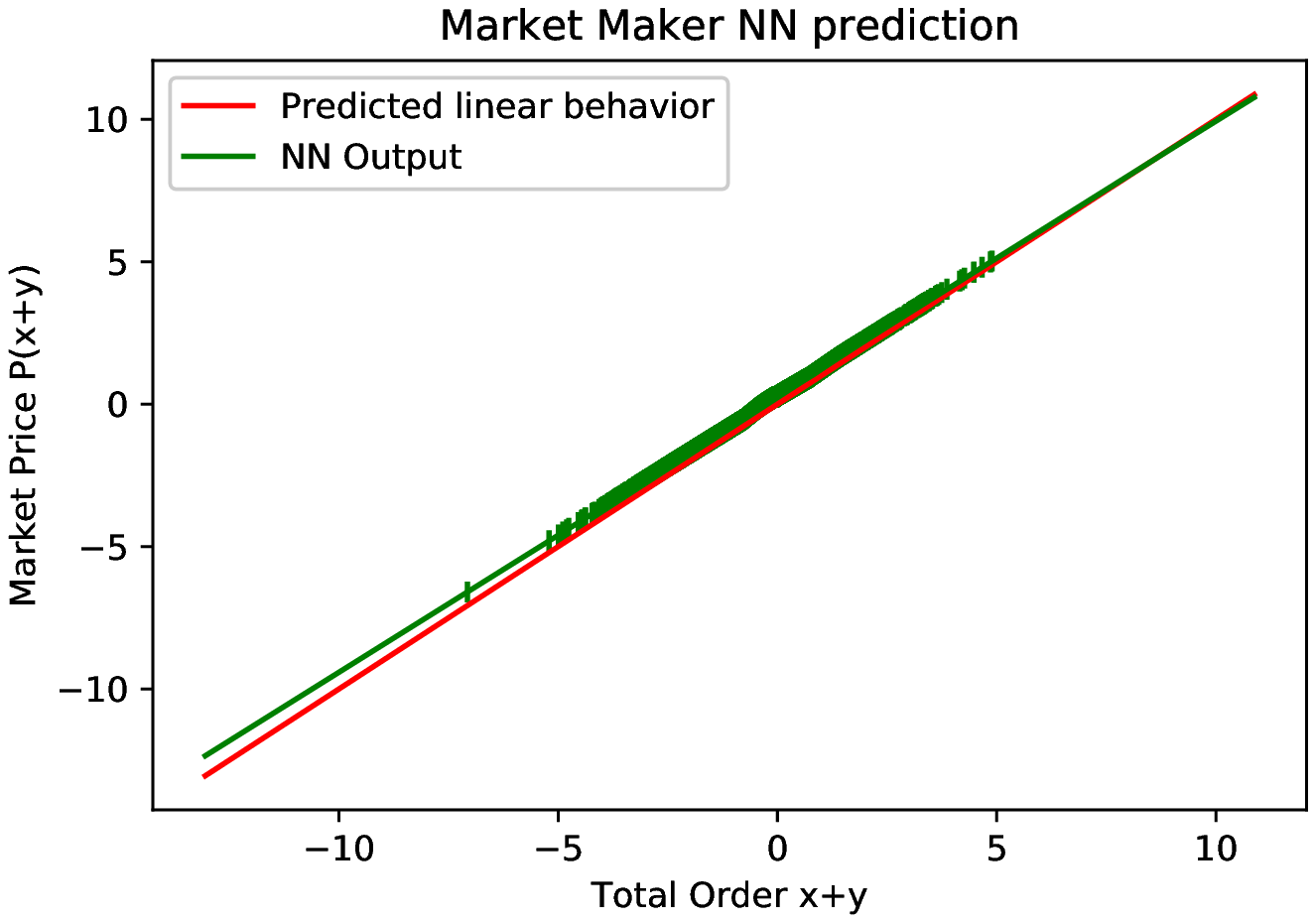}
 	  \label{fig:noisestart_cent_MM}
	\end{subfigure}
	
\caption{Assuming $\mu_z=0$, both neural networks learn the theoretical equilibrium when starting with a linear (top), approximately linear (middle) or Gaussian noise (bottom) insider order function.}
\label{fig:SP_plots}
\end{figure}
Figure \ref{fig:SP_plots} shows how both models converge nearly perfectly to the predicted linear equilibrium. The figure also shows the effects of using different initial order functions. Starting with an already linear (i.e. equilibrium fulfilling) insider function leads to fast convergence in around three loops while an approximately linear initial order function needs five loops to converge. However, the networks learn the equilibrium even if they start with an insider function that is indistinguishable from the noise trader and they do so within merely eight training loops. On our hardware, this training was bottlenecked by the CPU (likely due to the necessity of setting batch size $= 1$), so we opted to run it entirely on the CPU. A quad-core laptop CPU took around three to four minutes to reach convergence in the 5000 sample situation, and merely 30-60 seconds for 2000 samples.

\subsection{Altering the model, part I: Non-Gaussian distributions for $Y, Z$}

With our first goal, managing to train networks to learn Kyle's equilibrium, out of the way, we turn to uncharted territory and analyze the equilibrium by altering some key assumptions. We first drop the requirement of independent, normal distributions and later introduce transaction costs to the model. By using the same training regiment as above, we can reliably find the new equilibria in these situations that one could not theoretically derive otherwise.

So far, we have built and tested the model using various normal distributions to sample noise trader orders $Y$ and market prices $Z$. Normal distributions possess a defined mean and variance, they are symmetrical with one peak and no skewness. When dropping these assumptions, it is not at all clear how the system may behave. One key fact used in the theoretical derivation of the equilibrium is that the $L^2$-optimiser of the market maker's pricing rule is the least squares one. This stops being true when $Y$ or $Z$ are no longer normally distributed, as $X + Y$ is then no longer jointly normal (even if $X(Z)$ is linear and $Z$ or $Y$ respectively remain normal). By choosing different distributions for $Y$ and $Z$, we expect to see nonlinearities in the pricing and order functions that depend on the type of distributions and may correlate with how similar the used distributions are to normal ones in terms of their defining characteristics. 

\begin{figure}
\centering	
	\begin{subfigure}[b]{0.5\textwidth}
	\raggedleft
	   \includegraphics[width=0.75\linewidth]{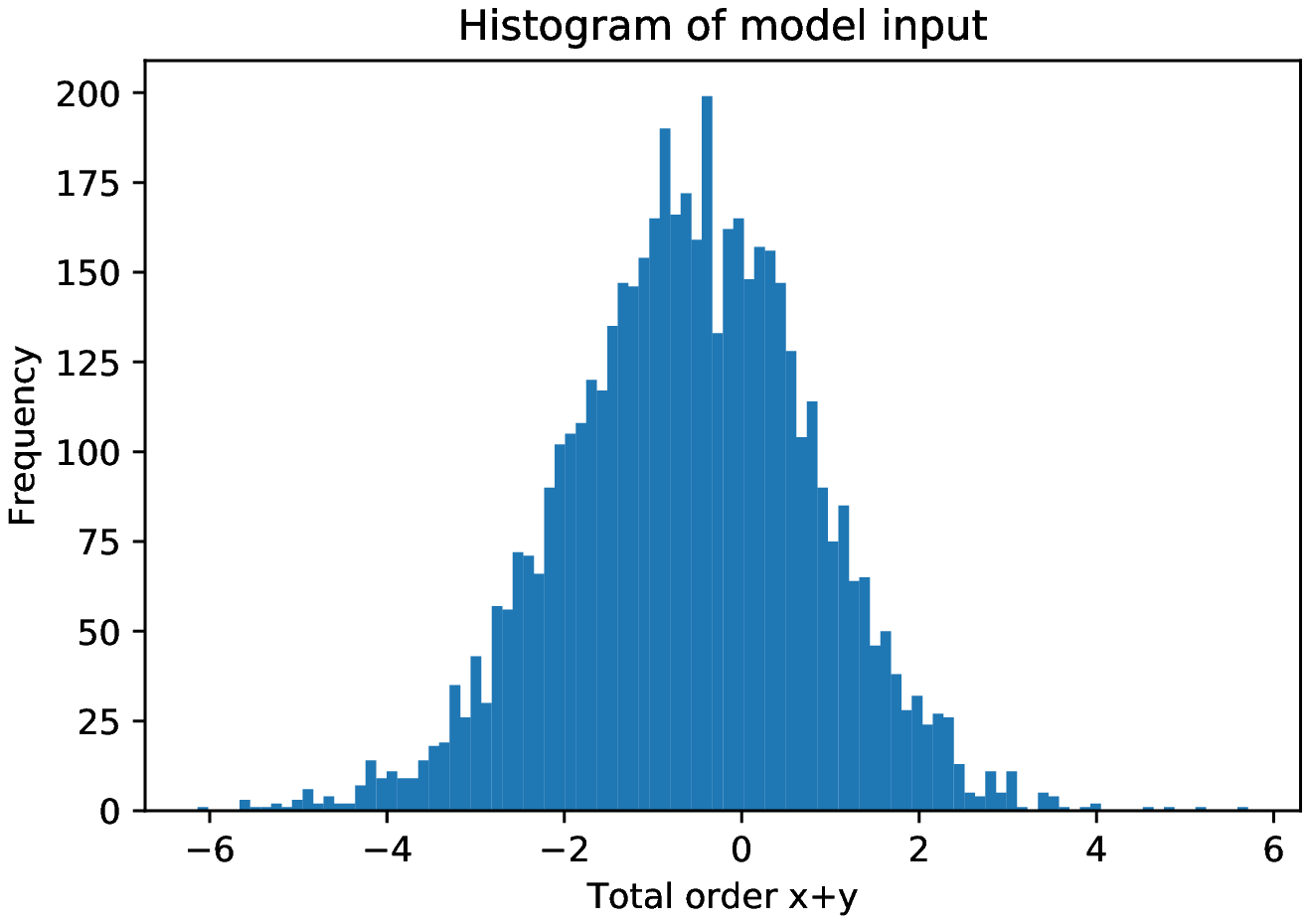}
 	   \label{fig:totord_laplace} 
	\end{subfigure}%
	\begin{subfigure}[b]{0.5\textwidth}
	\raggedright 
 	  \includegraphics[width=0.75\linewidth]{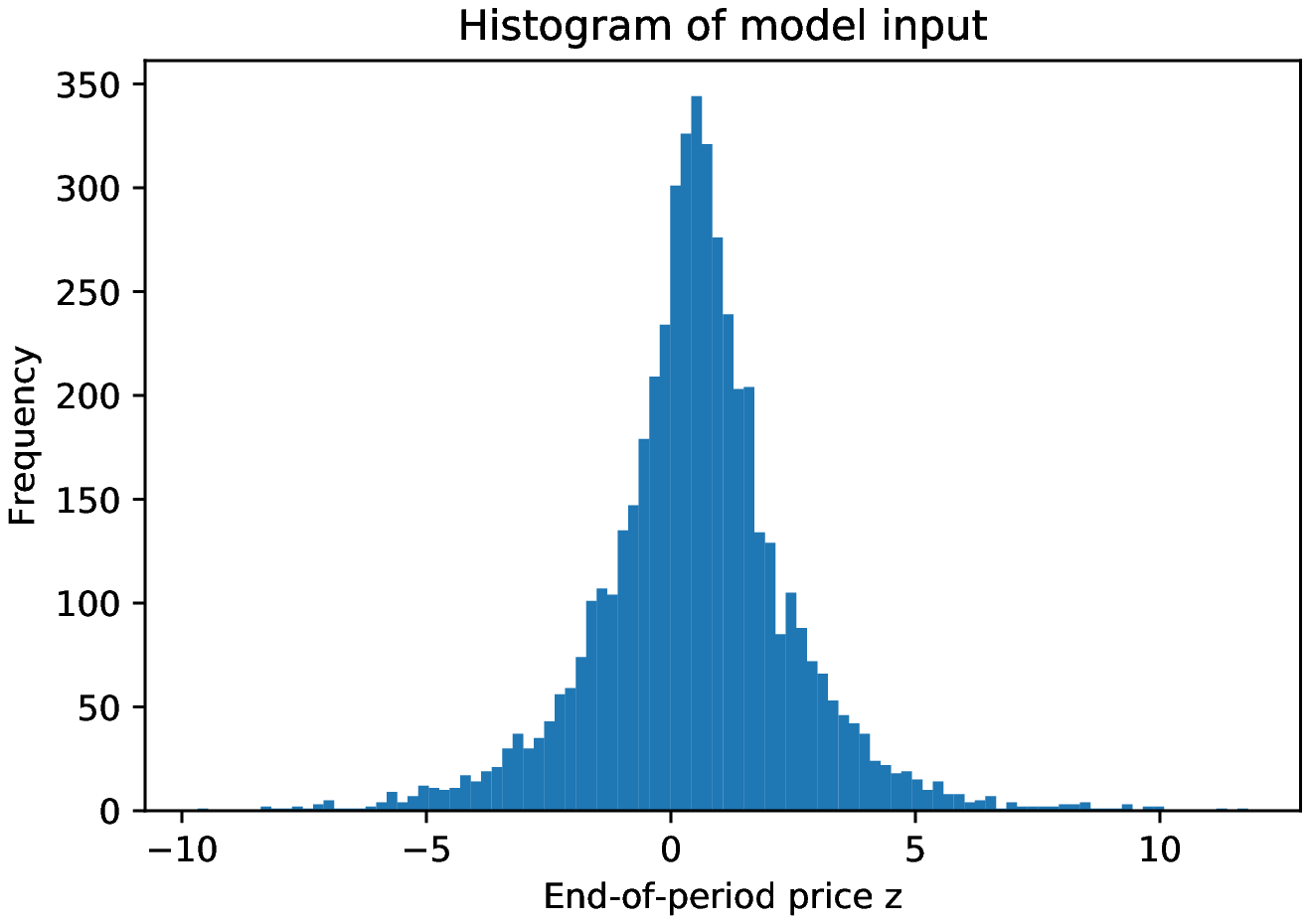}
 	  \label{fig:z_laplace}
	\end{subfigure} 
	
	\begin{subfigure}[b]{0.5\textwidth}
	\raggedleft
	   \includegraphics[width=0.75\linewidth]{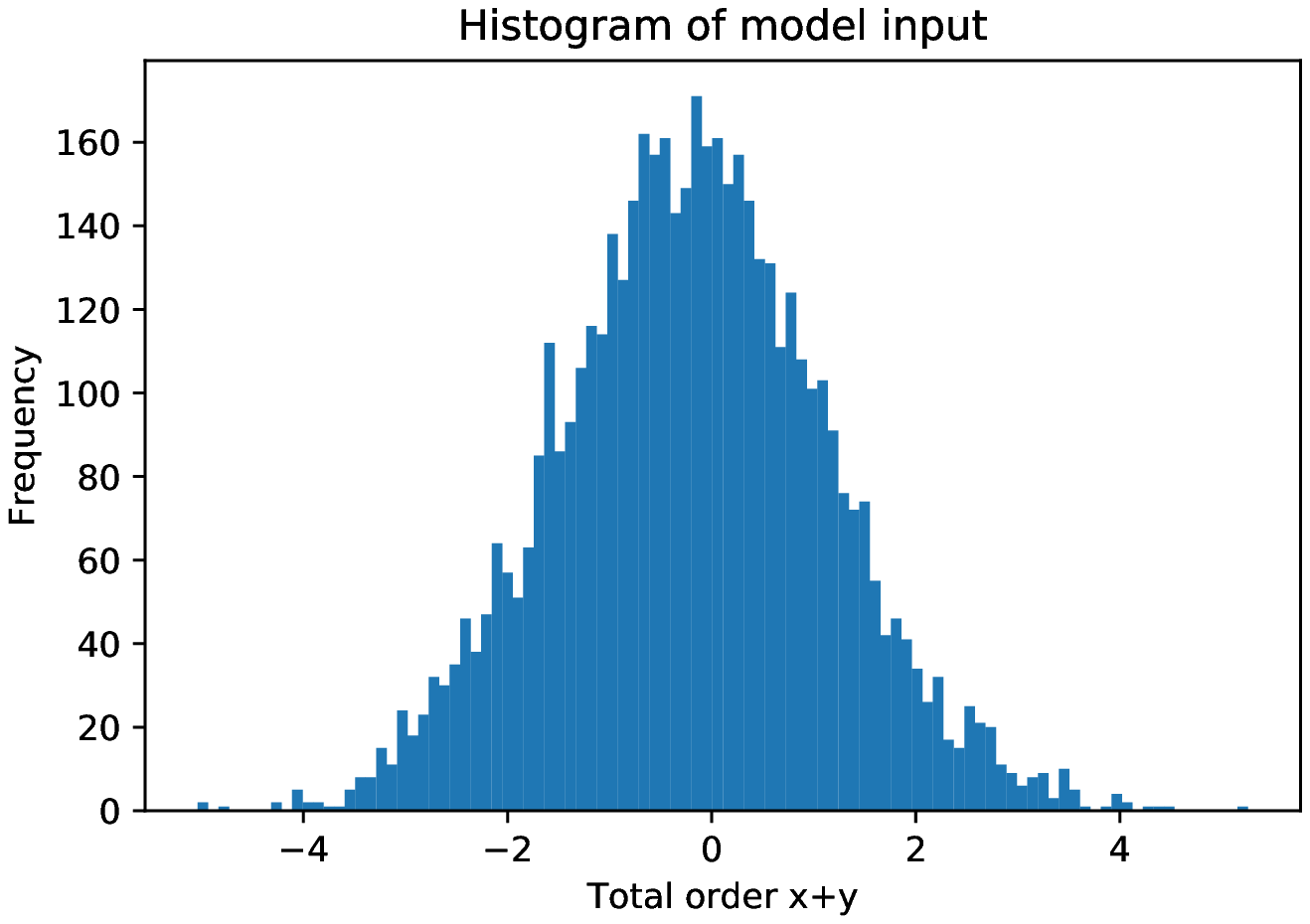}
 	   \label{fig:totord_gumbel} 
	\end{subfigure}%
	\begin{subfigure}[b]{0.5\textwidth}
	\raggedright 
 	  \includegraphics[width=0.75\linewidth]{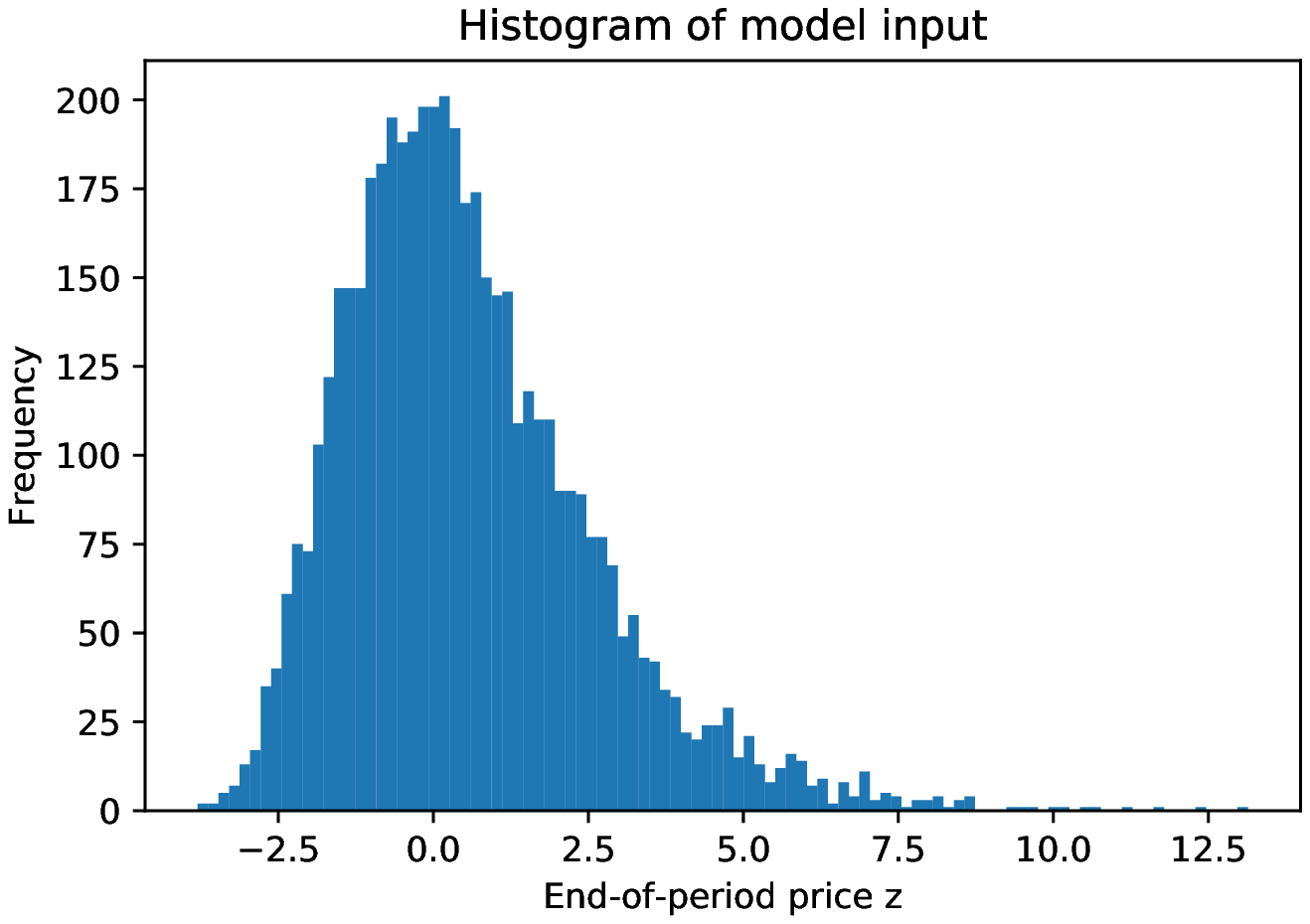}
 	  \label{fig:z_gumbel}
	\end{subfigure}
	
	\begin{subfigure}[b]{0.5\textwidth}
	\raggedleft
	   \includegraphics[width=0.75\linewidth]{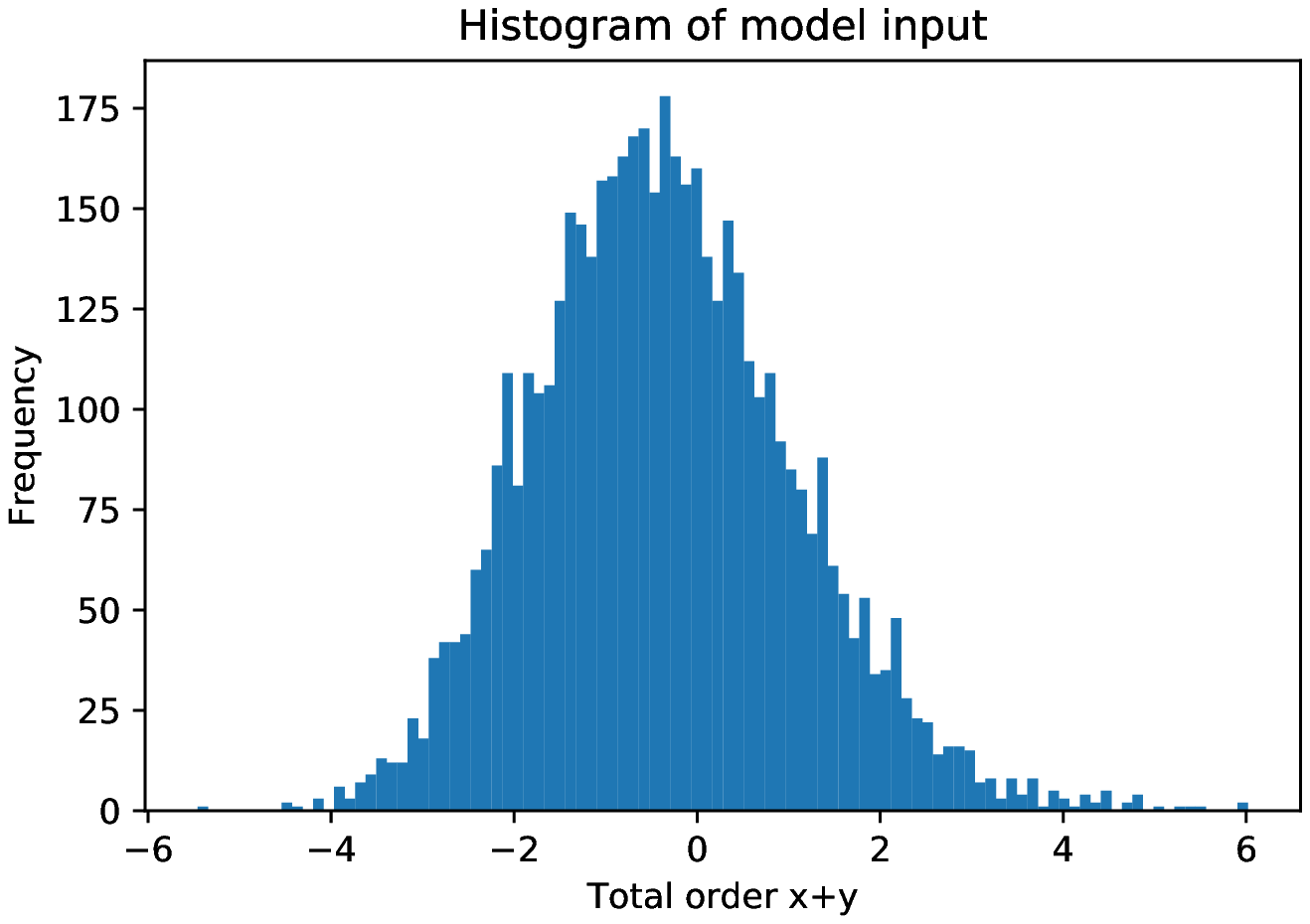}
 	   \label{fig:totord_centgamma} 
	\end{subfigure}%
	\begin{subfigure}[b]{0.5\textwidth}
	\raggedright 
 	  \includegraphics[width=0.75\linewidth]{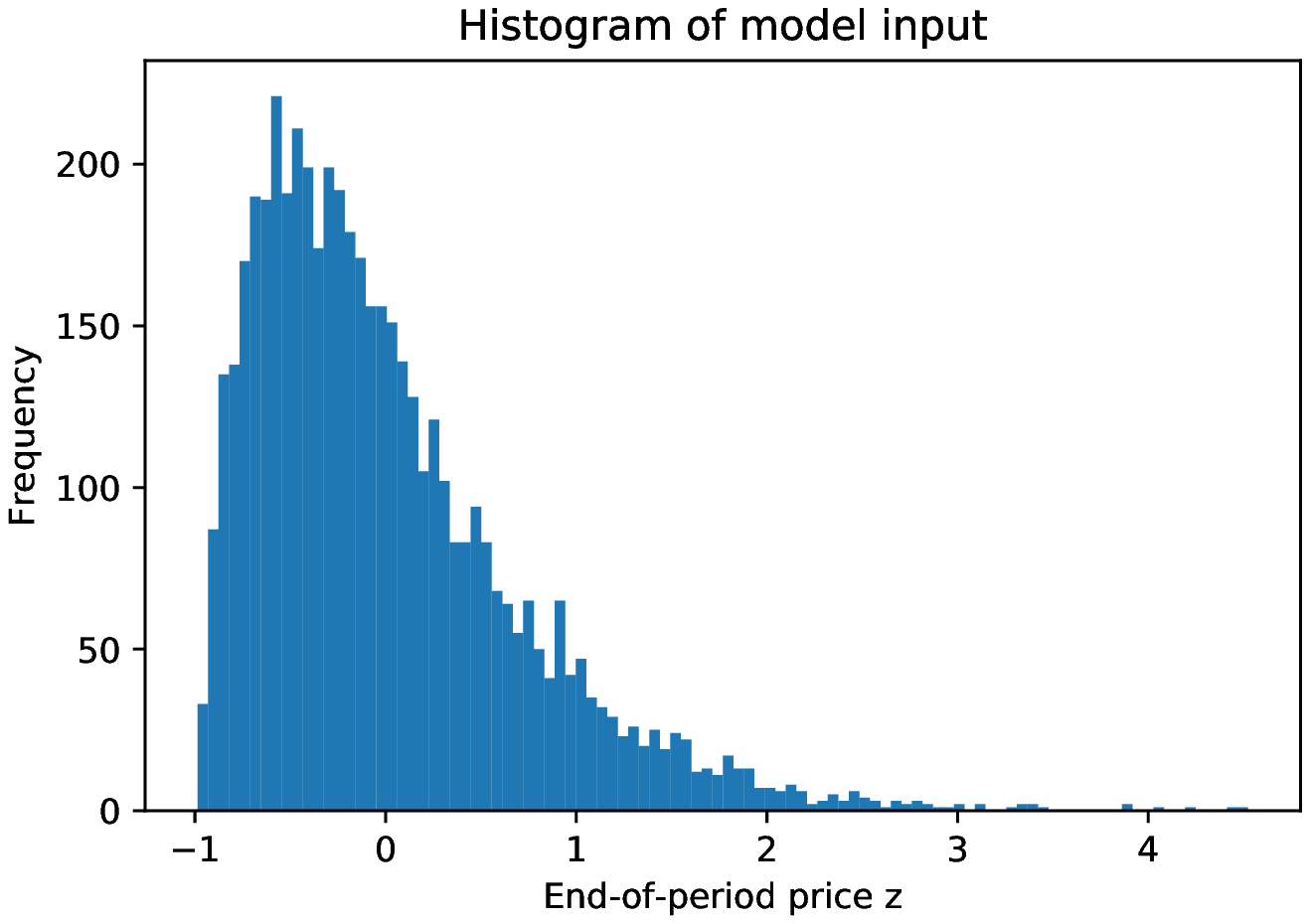}
 	  \label{fig:z_centgamma}
	\end{subfigure}
	
	\begin{subfigure}[b]{0.5\textwidth}
	\raggedleft
	   \includegraphics[width=0.75\linewidth]{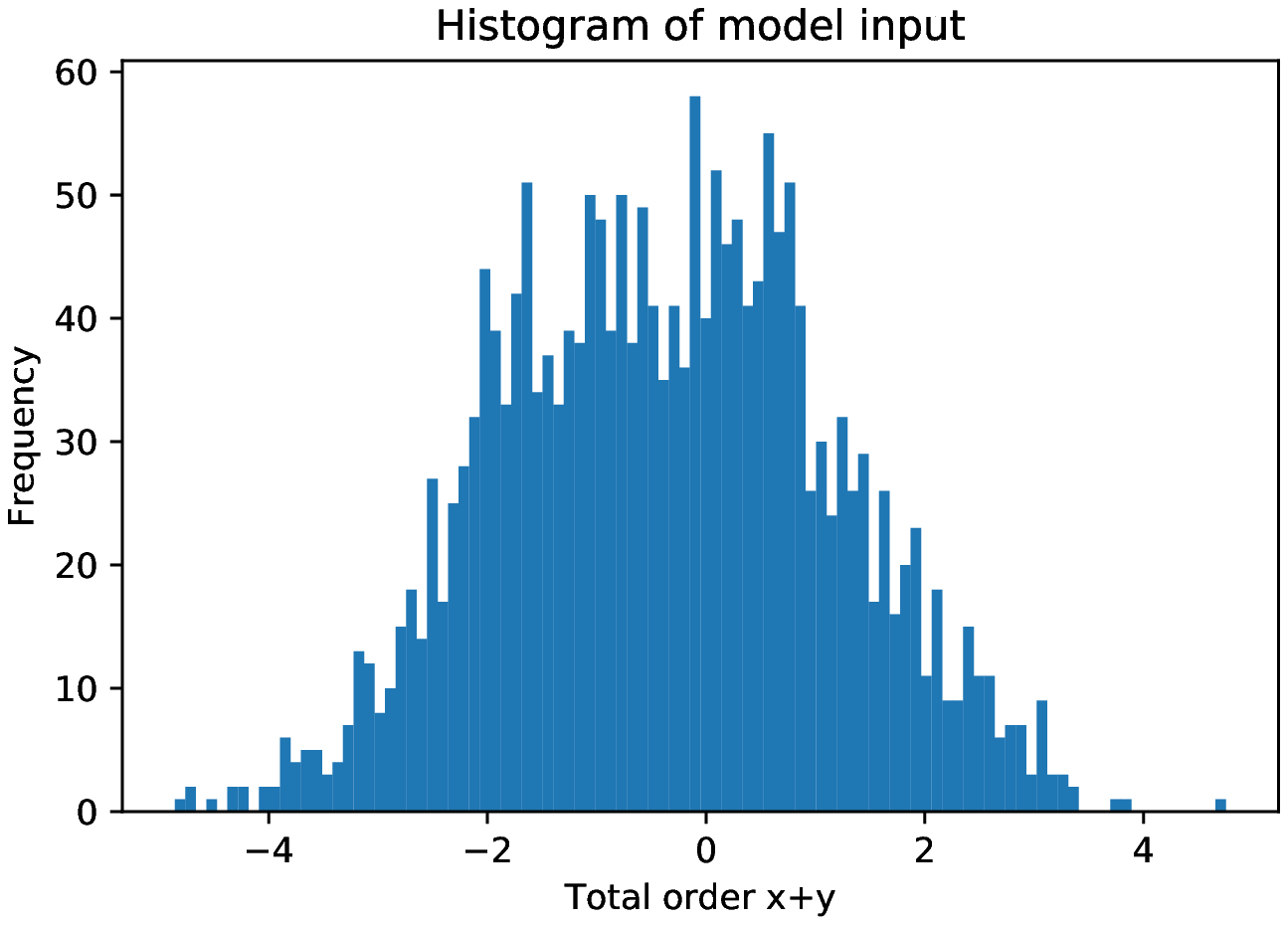}
 	   \label{fig:totord_bimodal} 
	\end{subfigure}%
	\begin{subfigure}[b]{0.5\textwidth}
	\raggedright 
 	  \includegraphics[width=0.75\linewidth]{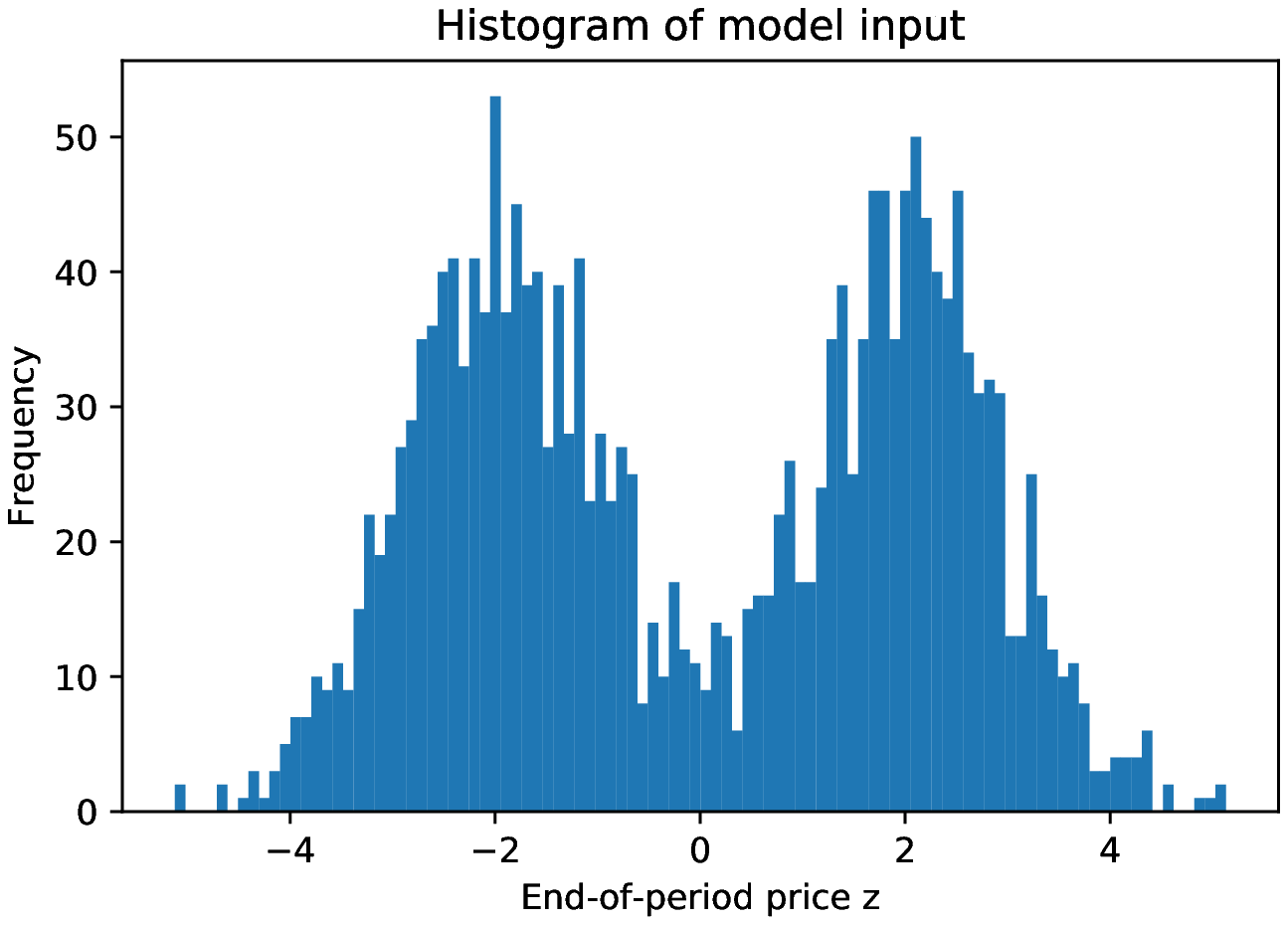}
 	  \label{fig:z_bimodal}
	\end{subfigure}
	
\caption{Comparing input distributions when picking different distribution families for $Z$. In order from top to bottom: Laplace, Gumbel, Gamma (centered), Bimodal. Plots on the right directly show the distribution of $Z$.}
\label{fig:nongauss_histograms}
\end{figure}

To that end, we picked several distributions with different attributes, namely the Laplace, Gumbel and Gamma distributions, as well as a bimodal distribution given by a mixture of two normal distributions. For the first three, mean and variance are defined and we thus selected their parameters for each distribution family such that mean and variance would match the ones chosen so far, i.e. $\E[Z]=\mu_z=0.5,\ \mathrm{Var}[Z]=\sigma_z^2=4,\ \E[Y]=\mu_y=0$ and $\mathrm{Var}[Y]=\sigma_y^2=1$. If $Y$ and $Z$ are chosen to both be distributed as Laplace, Gumbel or Bimodal, the model learns and outputs not only a linear equilibrium, but the one corresponding to the setting where $Y \sim \mathcal{N}(0,1)$ and $Z \sim \mathcal{N}(0.5,4)$. It is a somewhat surprising result, seeing as the Gumbel and Bimodal input distributions are recognisably different from the normal ones (see fig. \ref{fig:nongauss_histograms}). There are some slight nonlinearities present in the Bimodal predictions around the origin, but overall neural network predictions very much align with the linear optimum. 

\begin{figure}
\centering
	\begin{subfigure}[b]{0.5\textwidth}
	\raggedleft
	\includegraphics[width=0.9\linewidth]{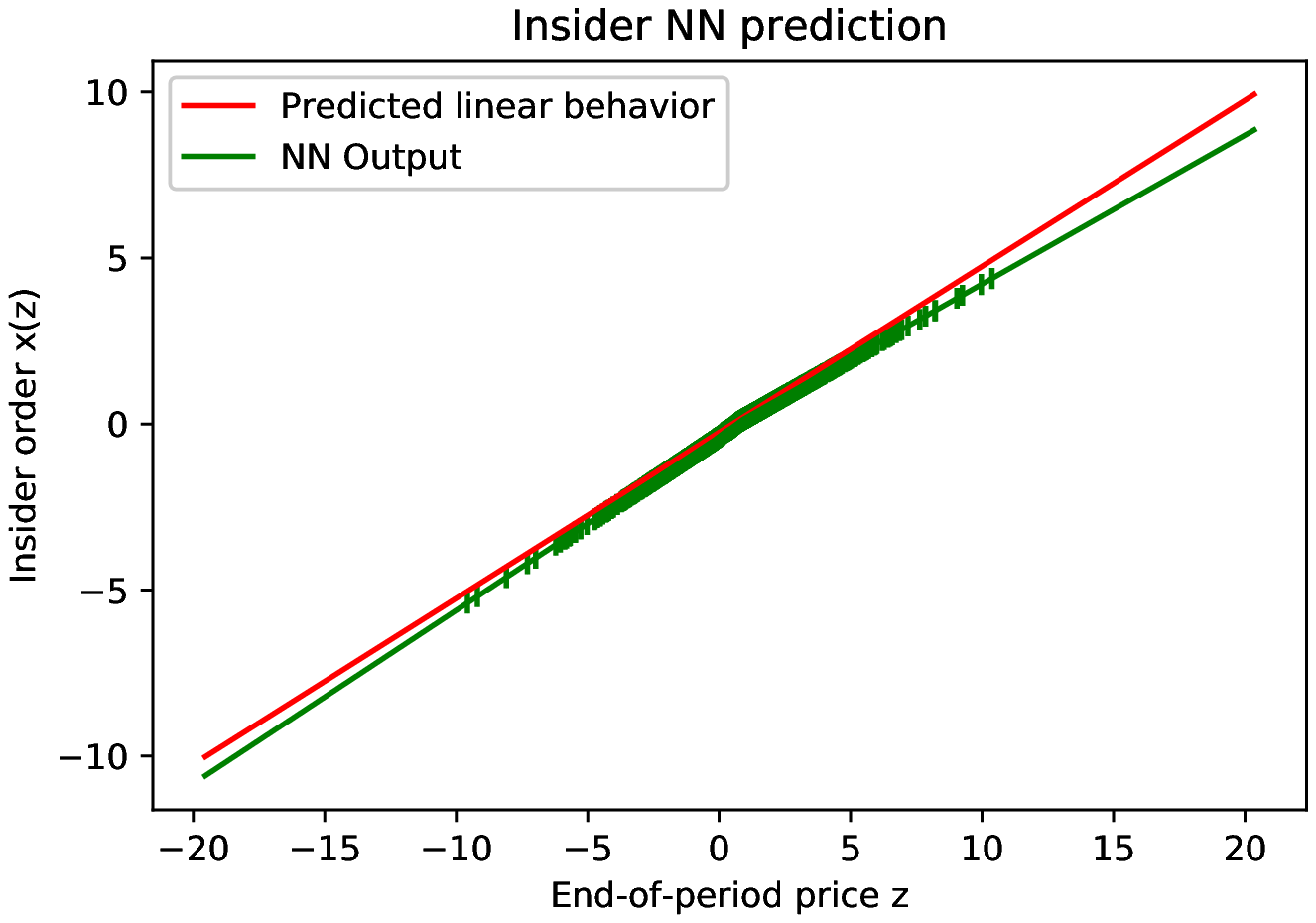}
	\label{fig:I_laplacen}
	\end{subfigure}%
	\begin{subfigure}[b]{0.5\textwidth}
	\raggedright 
	\includegraphics[width=0.9\linewidth]{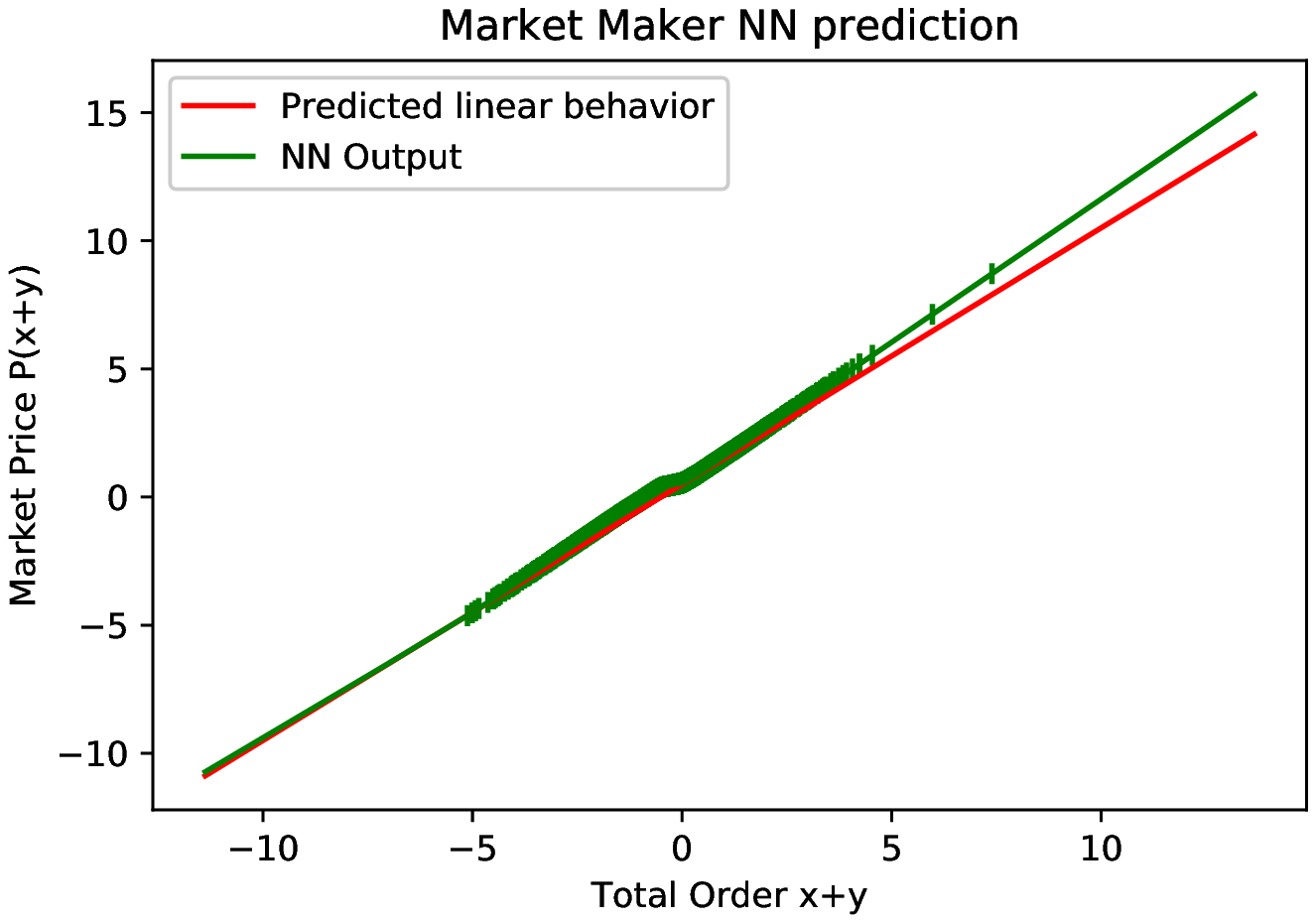}
	\label{fig:MM_laplacen}
	\end{subfigure}
	\begin{subfigure}[b]{0.5\textwidth}
	\raggedleft
	\includegraphics[width=0.9\linewidth]{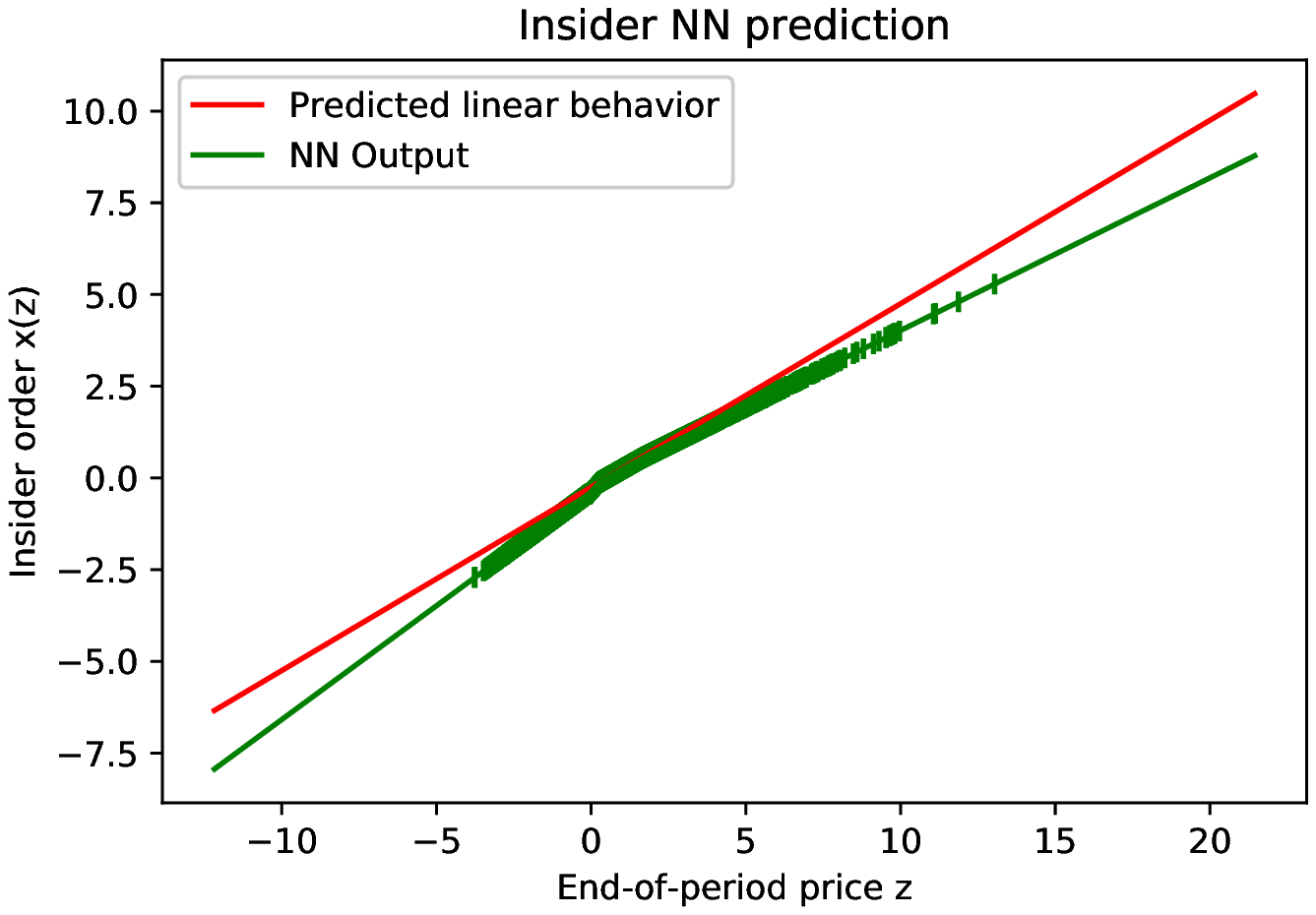}
	\label{fig:I_gumbeln}
	\end{subfigure}%
	\begin{subfigure}[b]{0.5\textwidth}
	\raggedright 
	\includegraphics[width=0.9\linewidth]{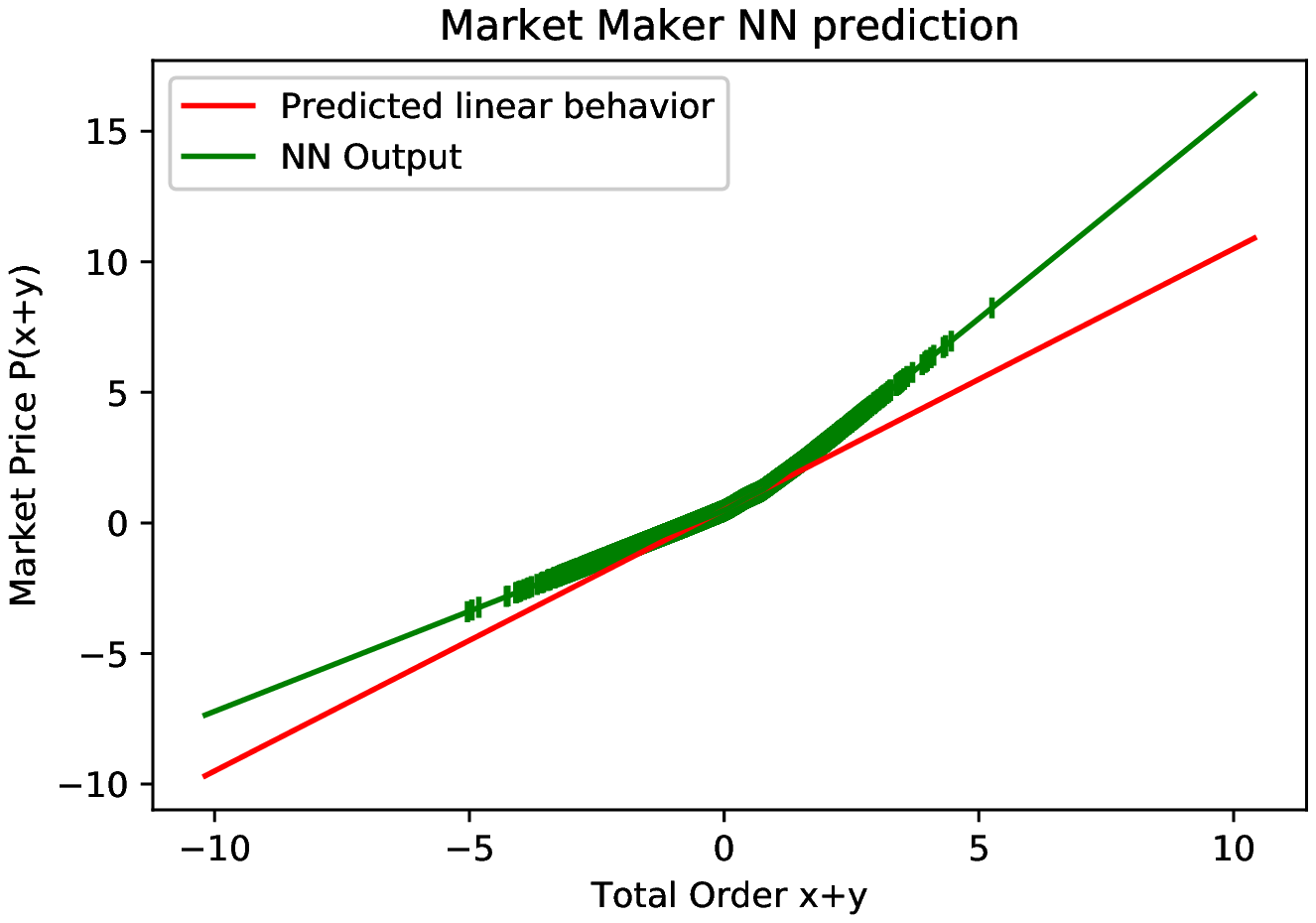}
	\label{fig:MM_gumbeln}
	\end{subfigure}
	\begin{subfigure}[b]{0.5\textwidth}
	\raggedleft
	   \includegraphics[width=0.9\linewidth]{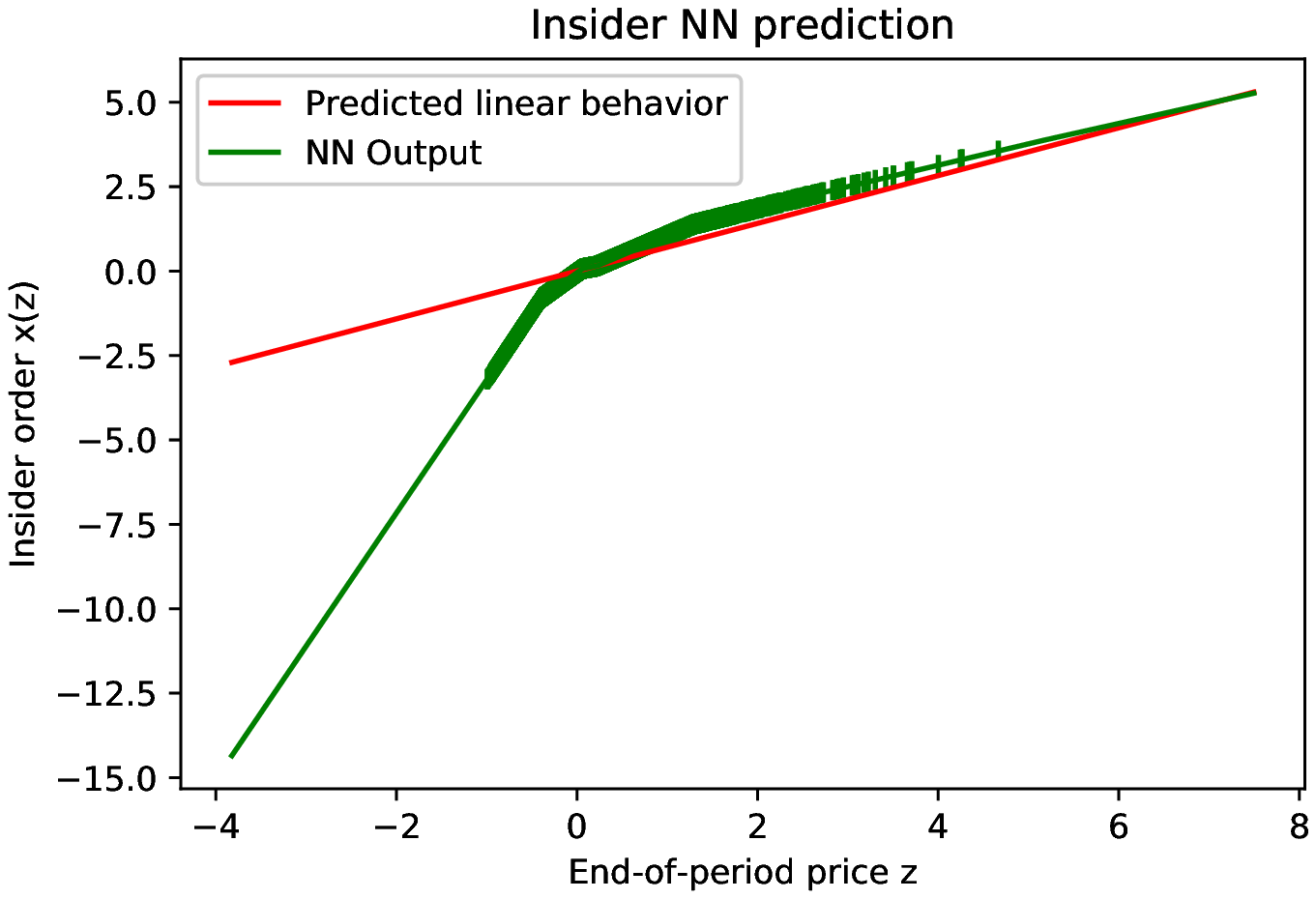}
 	   \label{fig:I_centgamman} 
	\end{subfigure}%
	\begin{subfigure}[b]{0.5\textwidth}
	\raggedright 
 	  \includegraphics[width=0.9\linewidth]{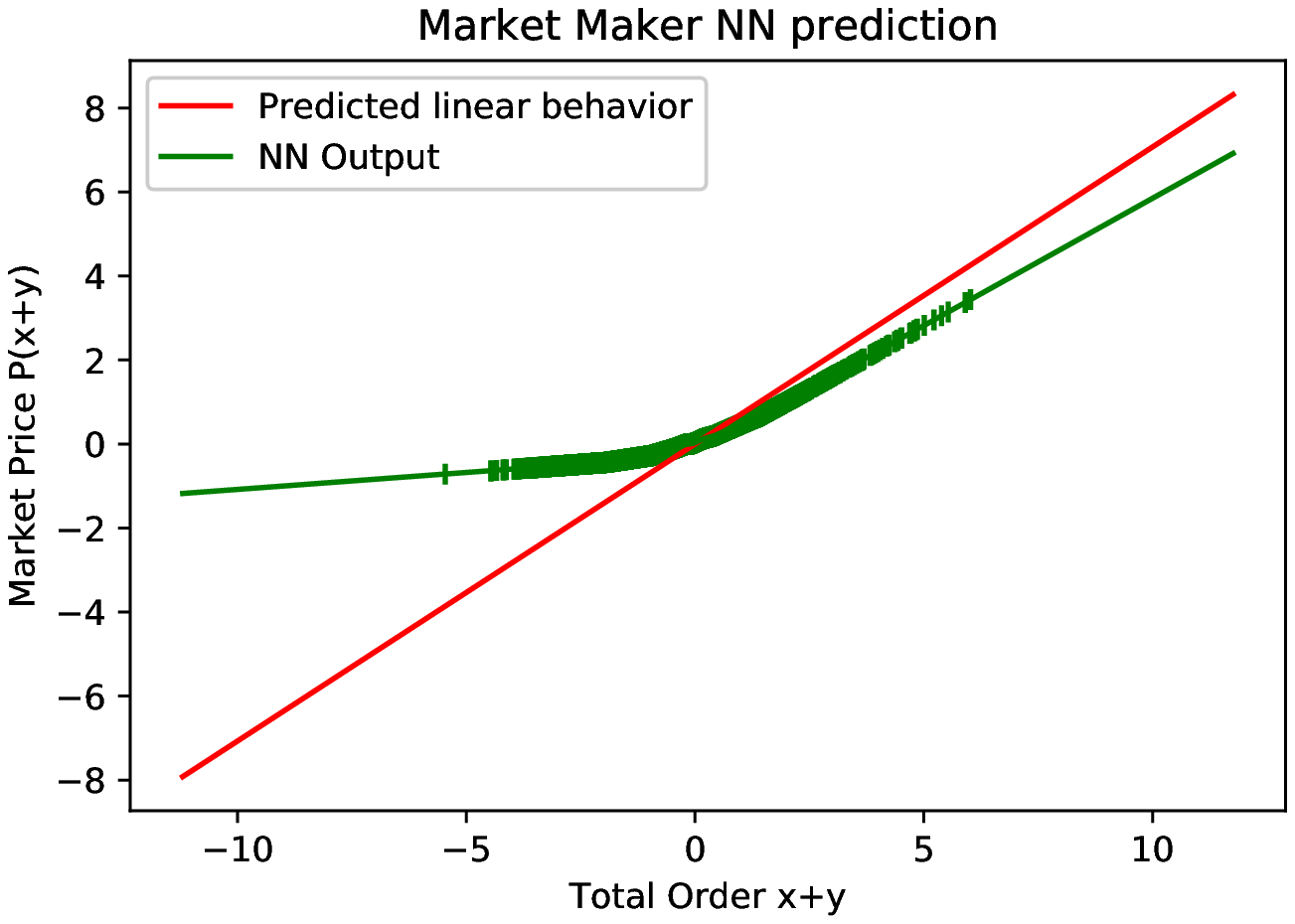}
 	  \label{fig:MM_centgamman}
	\end{subfigure} 
	\begin{subfigure}[b]{0.5\textwidth}
	\raggedleft
	   \includegraphics[width=0.9\linewidth]{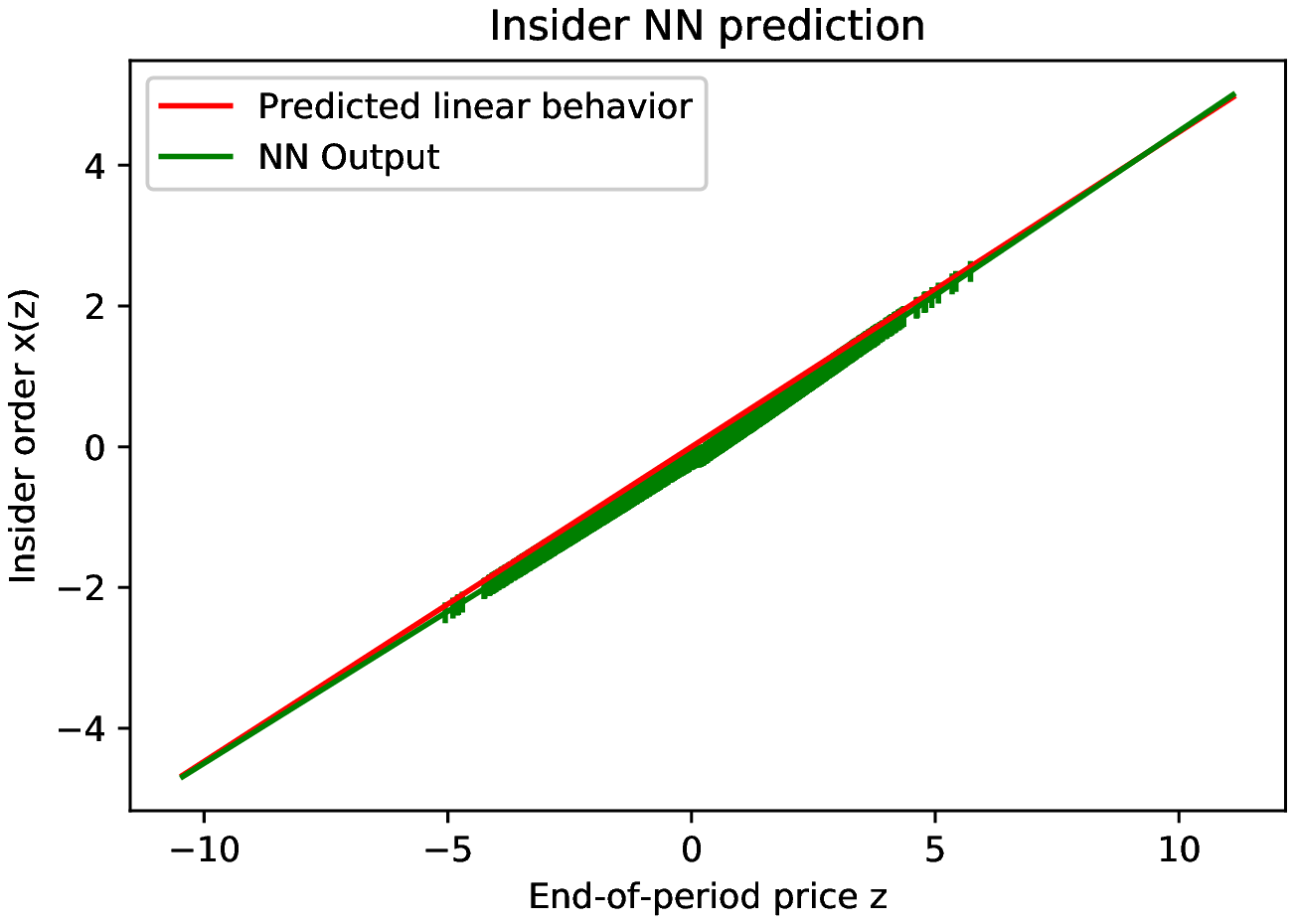}
 	   \label{fig:I_bimodaln} 
	\end{subfigure}%
	\begin{subfigure}[b]{0.5\textwidth}
	\raggedright 
 	  \includegraphics[width=0.9\linewidth]{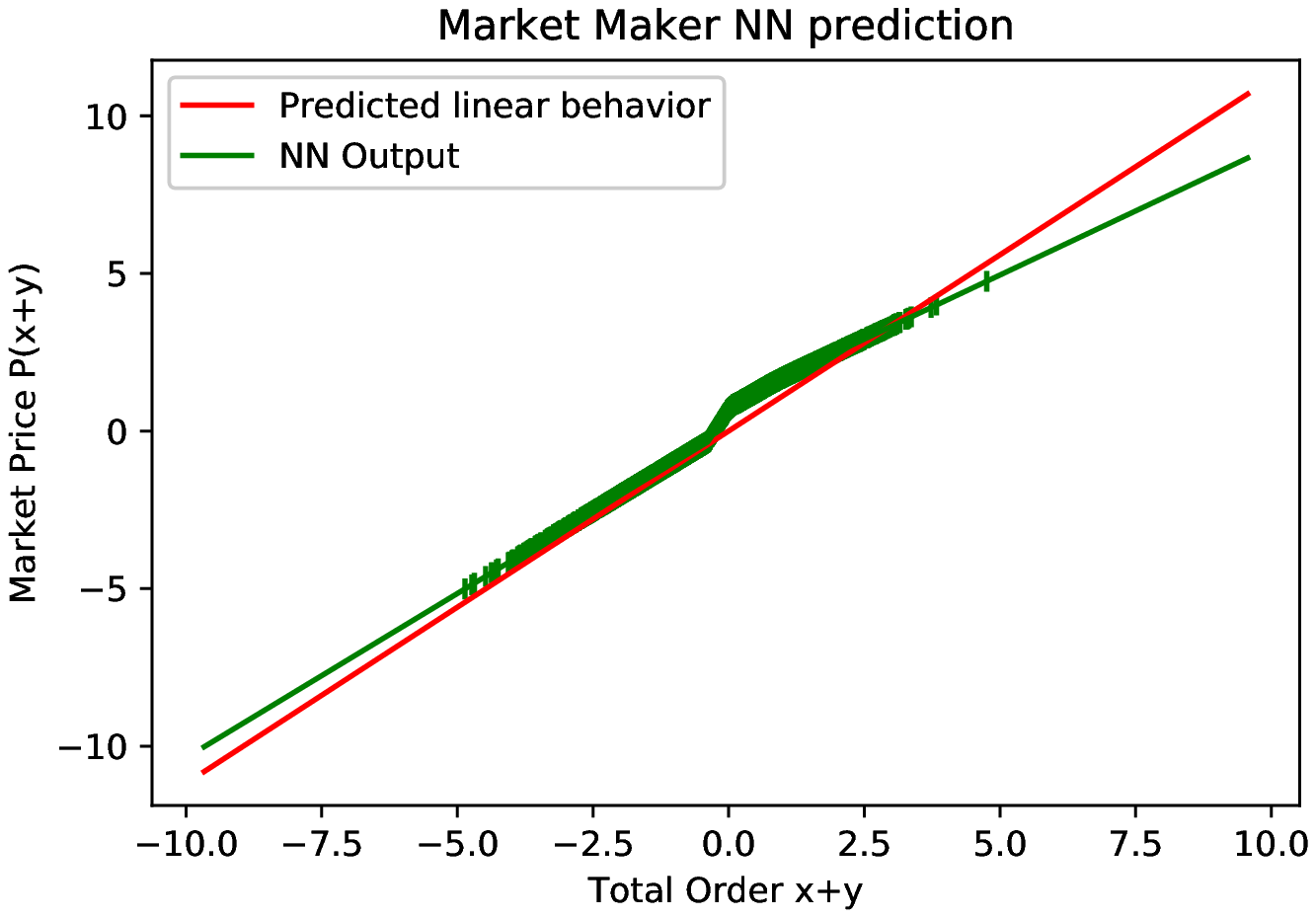}
 	  \label{fig:MM_bimodaln}
	\end{subfigure}
	
\caption{Model predictions when $Y$ is gaussian and $Z$ is sampled from different distributions. Top to bottom: Laplace, Gumbel, Gamma (centered) and Bimodal (normal mixture) distributions.}
\label{fig:nongauss_diffdists_predictions}
\end{figure}

A second step consisted of leaving $Y$ normally distributed as $Y\sim\mathcal{N}(0,1)$ while altering the distribution of $Z$. We show results for Laplace, Gumbel, Gamma and Bimodal distributions in Figure \ref{fig:nongauss_diffdists_predictions} while Figure \ref{fig:nongauss_histograms} shows the distributions of the resulting model input data as histograms. The results are more mixed now. When $Z$ is sampled from Laplace or Bimodal distributions, we observe the same linear equilibrium as in the normal situation. If Z is sampled from a Gumbel distribution, both pricing and order functions are linear for positive and negative values with a bend around the origin. For each function, its two slopes are similar to the one predicted in a normal model. These significant, but not major differences to the normally distributed situation correspond well to the comparison between a Gumbel and normal distribution: similar overall look, but small deciding differences (skewness). When sampling $Z$ from a Gamma distribution, we thus expect to see larger differences to the normal situation.

We have previously seen that the model struggles with data that shows exclusively positive values. We thus alter the parameters of the Gamma distribution to correspond to a mean of 1 and variance of 2, which produces a sample of $Z$ with values mostly between 0 and 2 and a significant tail. We then subtract 1 from that sample. The distribution is thus centered, but still with a large skew due to being a Gamma distribution. We expect a result similar to the Gumbel one, although the skewness is much larger here. We compare the predictions to the equilibrium situation. Both function predictions are again piecewise linear with a single bend around the origin. Both show a large deviation for negative values while having a similar (pricing function) or nearly identical (order function) slope as the normal theoretical optimum. This result further indicates that one, the model is very sensitive to non-centered data and two, skewness leads to piecewise linear predictions, where the amount of skewness influences how closely slopes align with the normal situation.

\subsection{Altering the model, part II: Introducing transaction costs}

We introduce market frictions to the model. Concretely, we assume that the insider now pays a transaction fee proportional to their order size. This is implemented by adding a penalisation term to the insider's loss function. The new version works as follows:

\begin{itemize}
	\item Sample $z \sim Z$ (as model input) and $y \sim Y$.
	\item Calculate a loss for each pair $(z_i,y_i)$:
	$$L(z_i,y_i):= -[(z-M(I(z)+y))I(z) - \epsilon \cdot |I(z)|]$$
	\item The total loss over one epoch of training is then given by:
	$$L_\text{epoch} = -\E_{(Z,Y)}[(Z-M(I(Z)+Y))I(Z) - \epsilon \cdot |I(Z)|].$$
\end{itemize}

Epsilon represents the fraction of the total order size that must be paid as transaction cost. This penalty only impacts the inside trader, so we do not expect the market maker's behaviour to change considerably. The inside trader, however, should adjust their behaviour. One could expect them to not trade at all in a certain price range where transaction costs would be larger than the expected payoff. If we assume that the market maker stays unimpacted (i.e. uses Kyle's predicted optimal pricing function), then the boundary of that price range would lie wherever the expected payoff is zero. Looking at the above loss function, for a given z this happens when:
\begin{align*}
 0 &= \E[(z-P(x(z)+Y))x(z) - \epsilon \cdot |x(z)|]  \\
\Leftrightarrow \epsilon &= \pm \left(z-\frac{\sigma_z}{2\sigma_y} x(z) - \mu_z\right) \\
\Leftrightarrow x(z) &= \frac{2\sigma_y}{\sigma_z} \cdot (z - (\mu_z \pm \epsilon))
\end{align*}

On the other hand, the range where $x(z)=0$ holds should then be bounded by the $z$ for which $z = \mu_z \pm \epsilon$. In our non-centered parameter setting, this corresponds to $z = 0.5 \pm \epsilon$. \\

\begin{figure}
\centering
	\begin{subfigure}[b]{0.33\textwidth}
	\centering
	\includegraphics[width=0.95\linewidth]{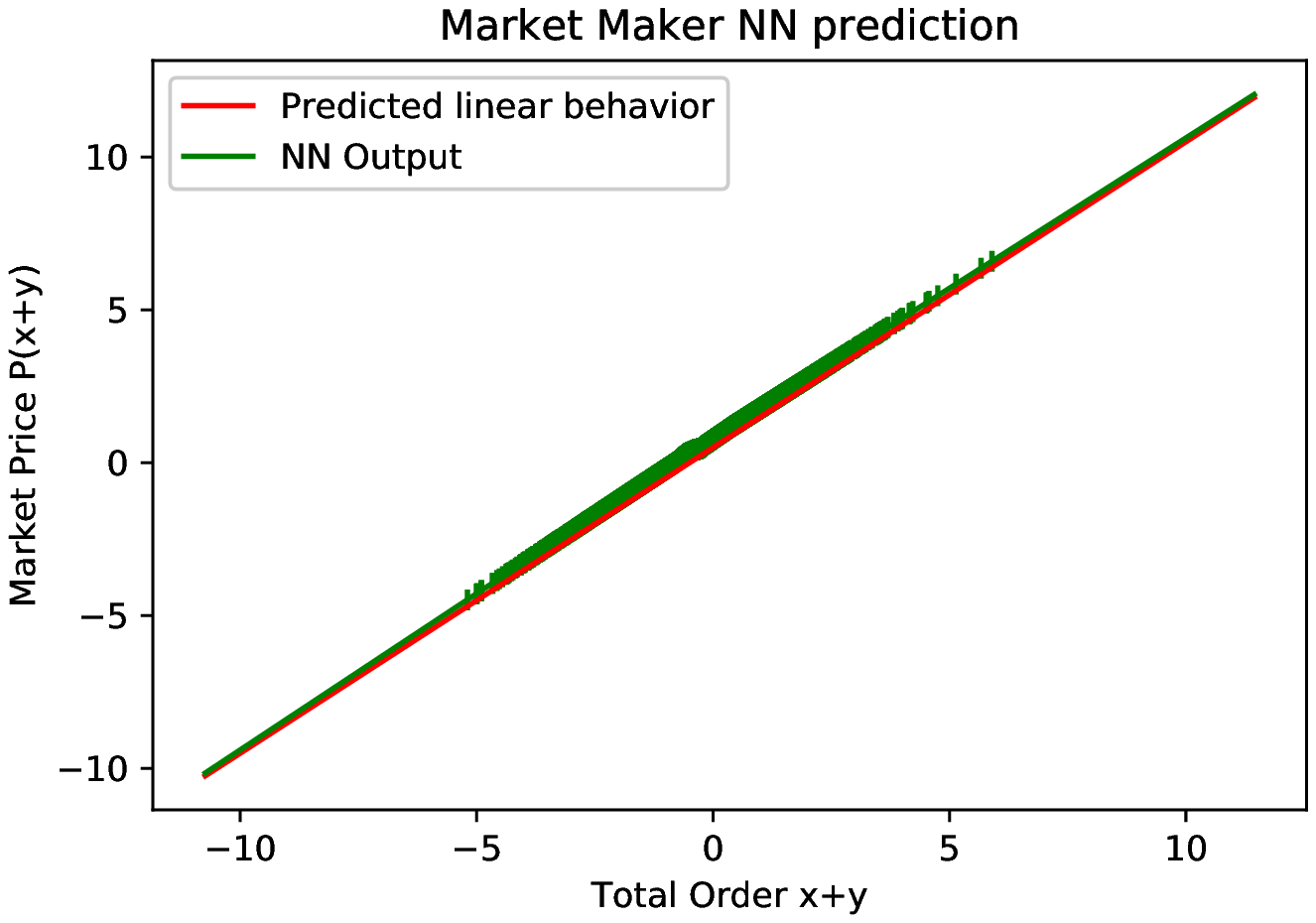}
	\label{fig:MM_01}
	\end{subfigure}%
	\begin{subfigure}[b]{0.33\textwidth}
	\centering
	\includegraphics[width=0.95\linewidth]{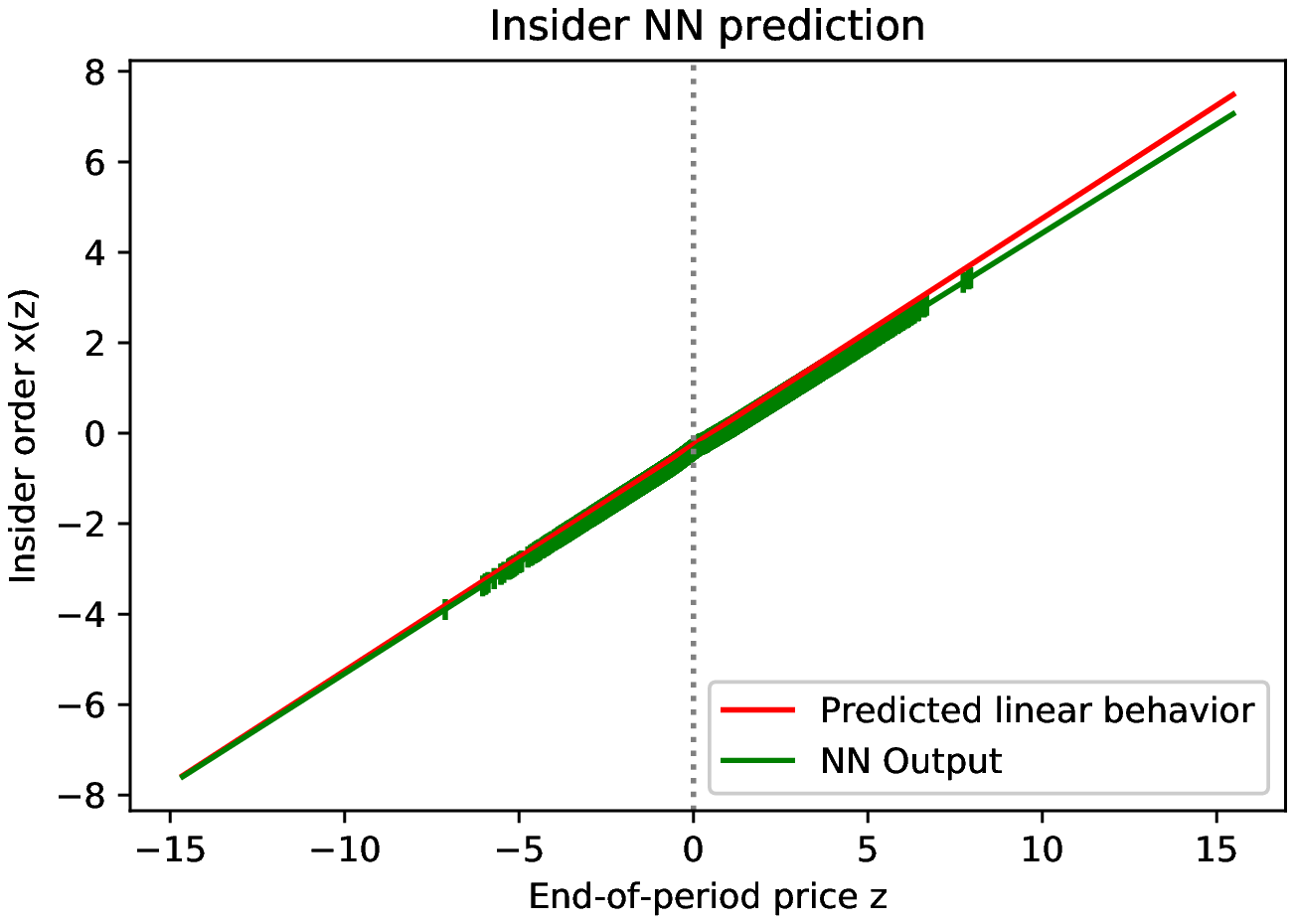}
	\label{fig:I_01_preds}
	\end{subfigure}%
	\begin{subfigure}[b]{0.33\textwidth}
	\centering
	\includegraphics[width=0.95\linewidth]{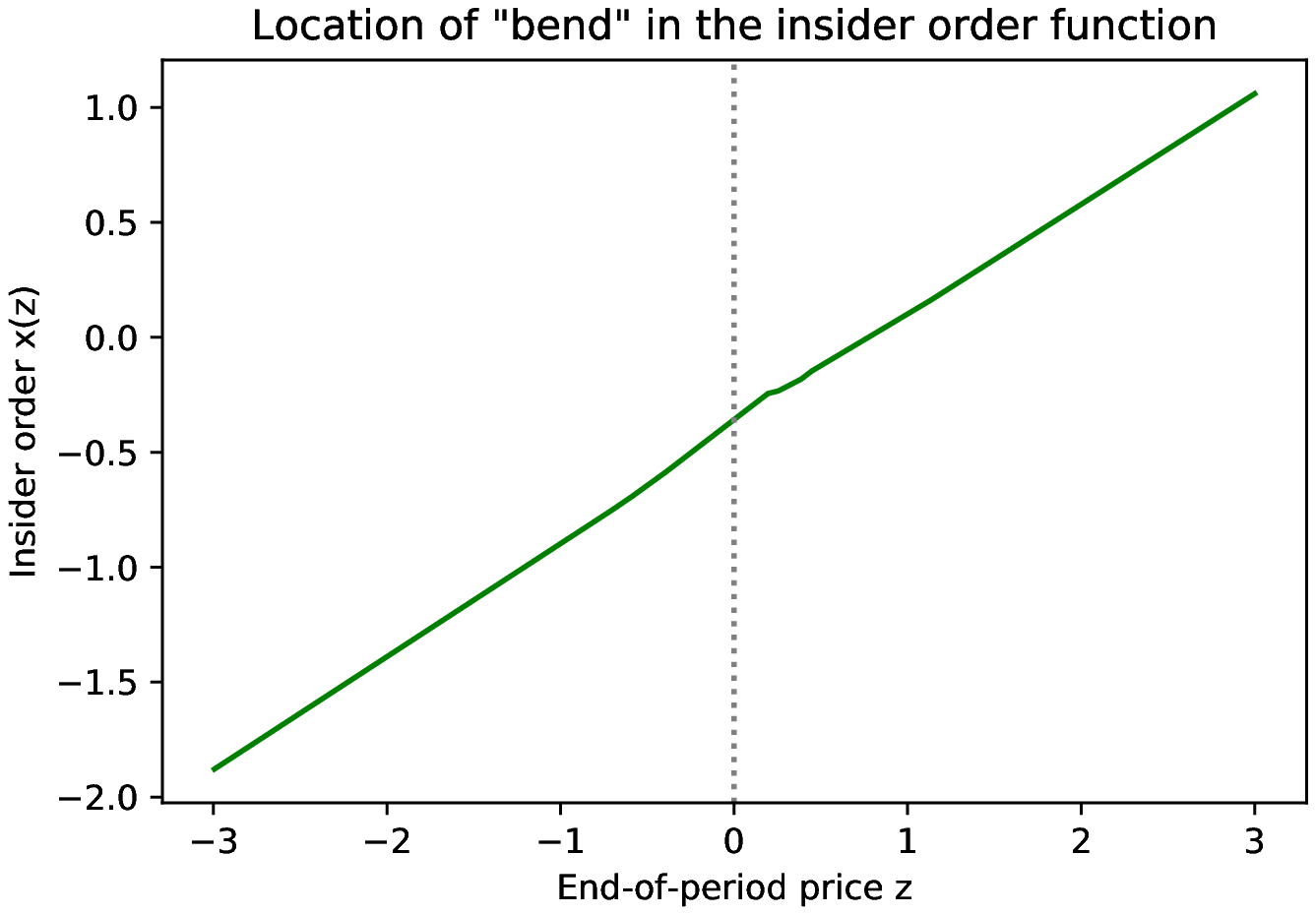}
	\label{fig:I_01_bends}
	\end{subfigure}
	
	\begin{subfigure}[b]{0.33\textwidth}
	\centering
	\includegraphics[width=0.95\linewidth]{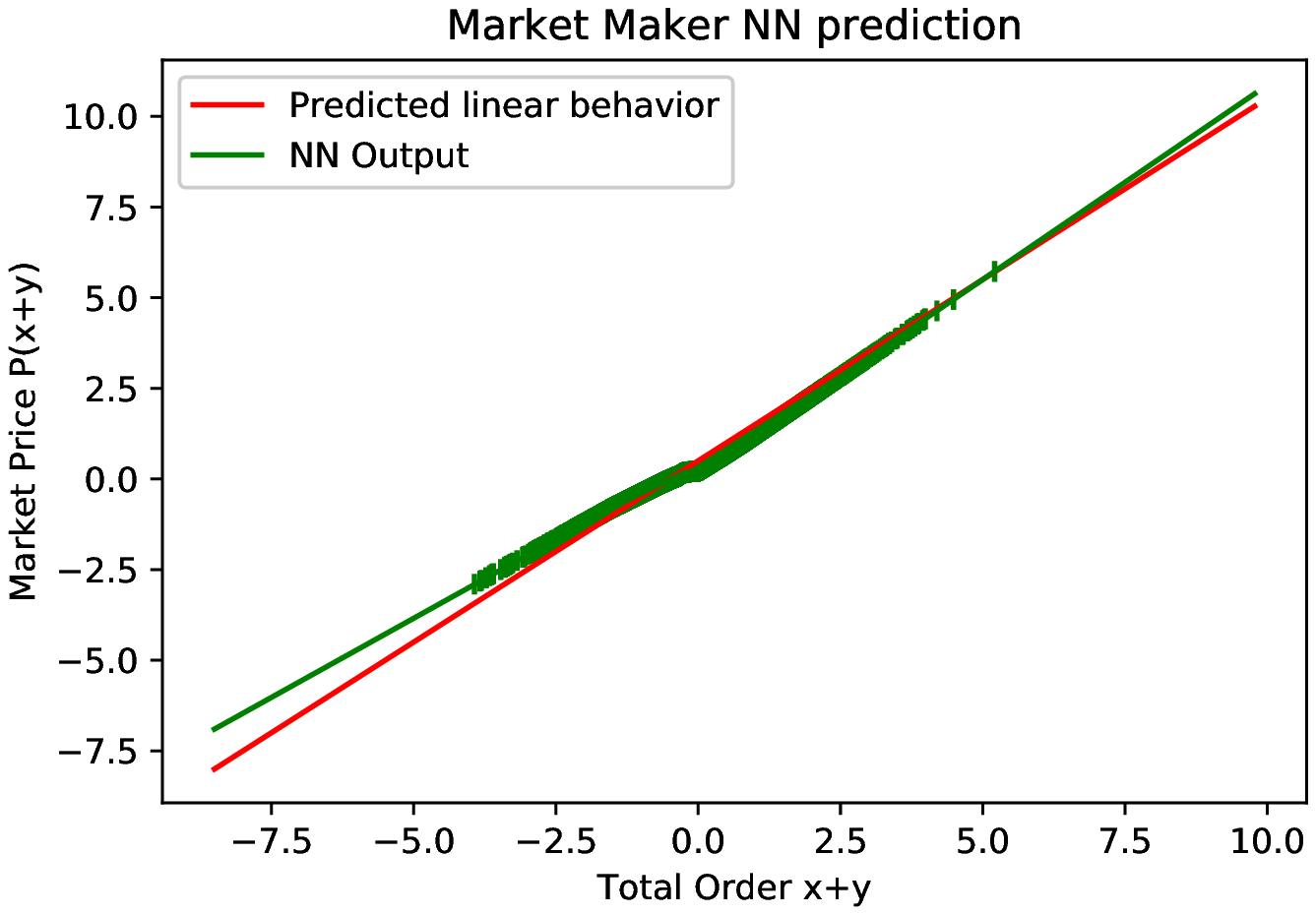}
	\label{fig:MM_075}
	\end{subfigure}%
	\begin{subfigure}[b]{0.33\textwidth}
	\centering
	\includegraphics[width=0.95\linewidth]{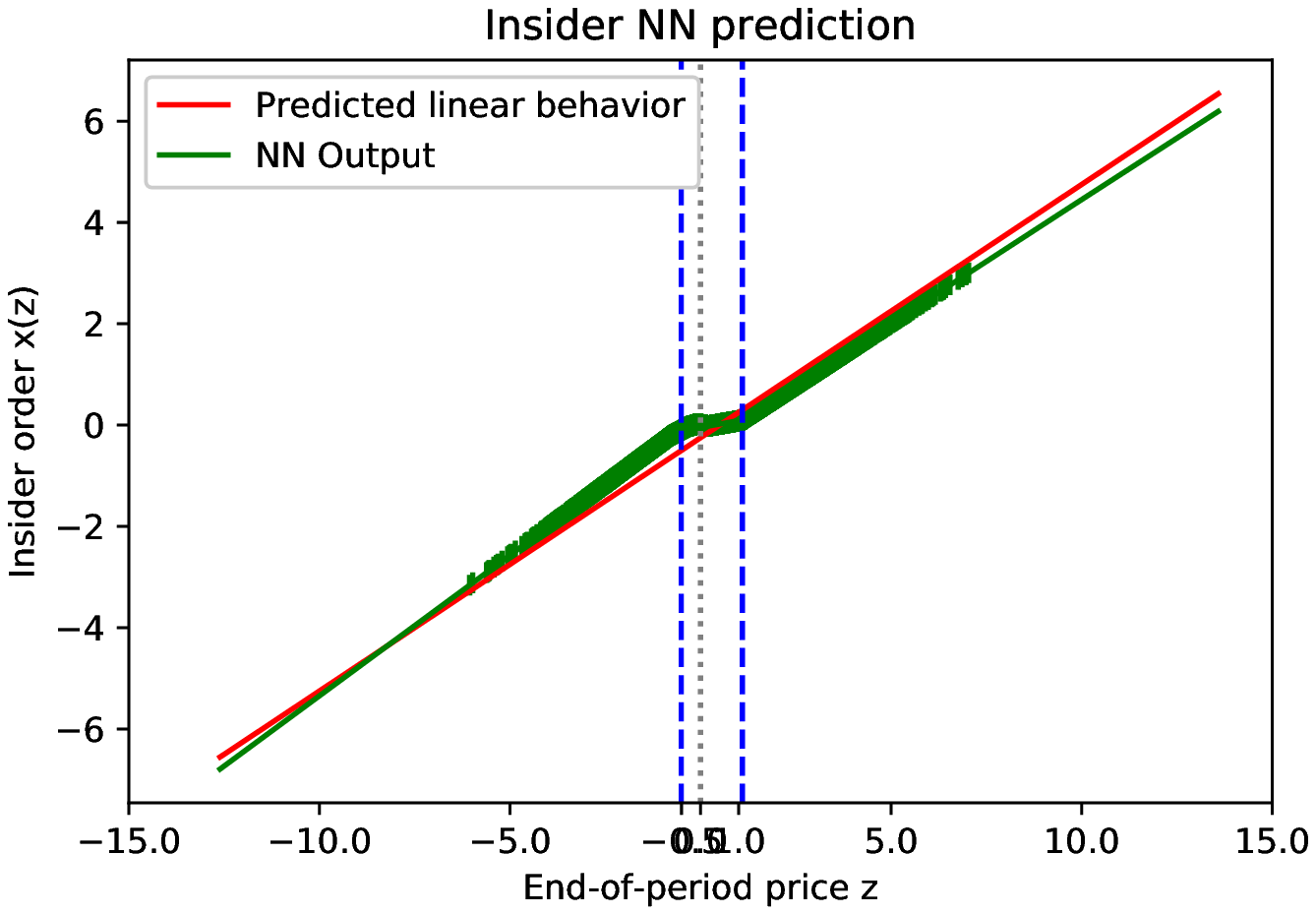}
	\label{fig:I_075_preds}
	\end{subfigure}%
	\begin{subfigure}[b]{0.33\textwidth}
	\centering
	\includegraphics[width=0.95\linewidth]{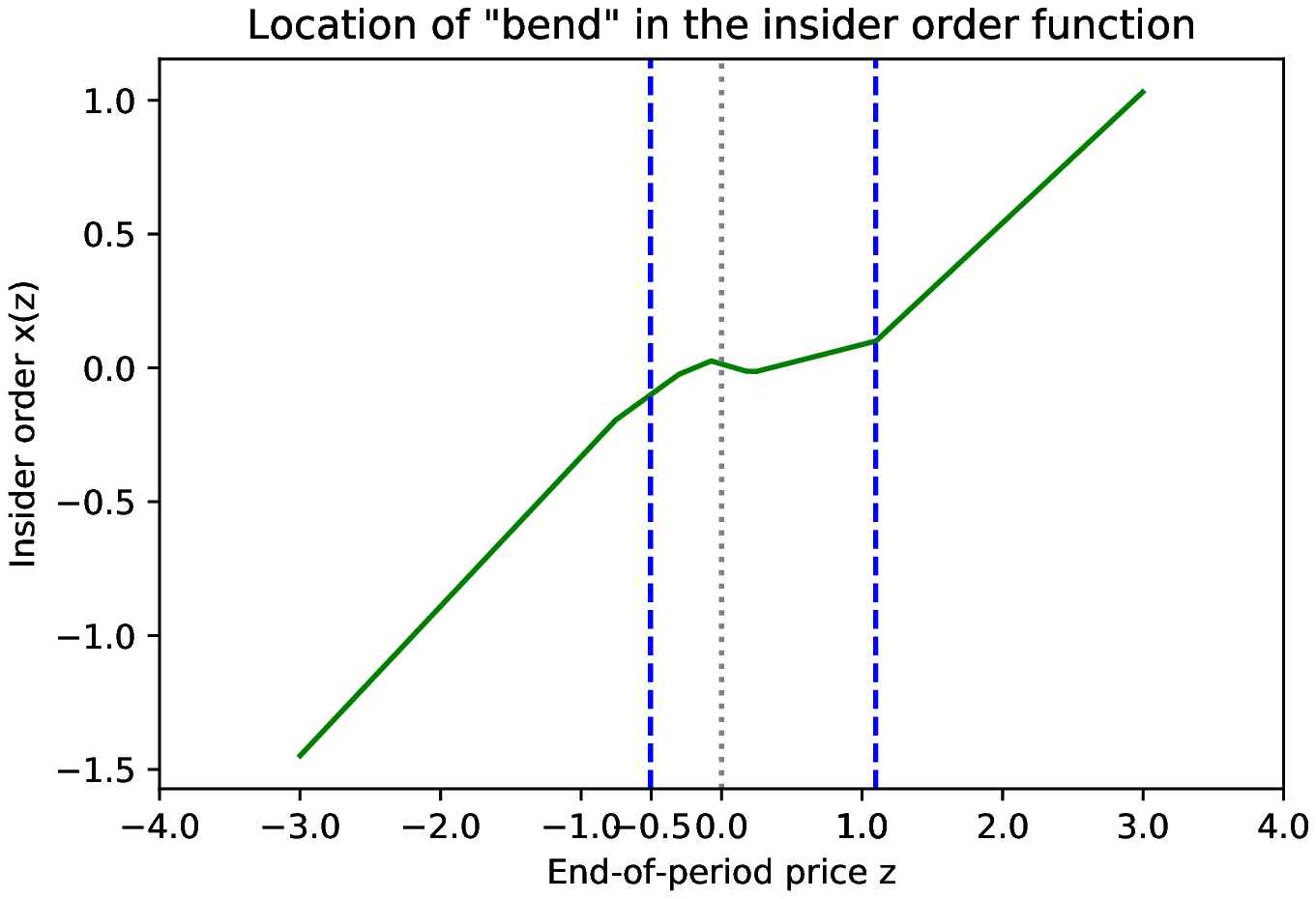}
	\label{fig:I_075_bends}
	\end{subfigure}
	
	\begin{subfigure}[b]{0.33\textwidth}
	\centering
	\includegraphics[width=0.95\linewidth]{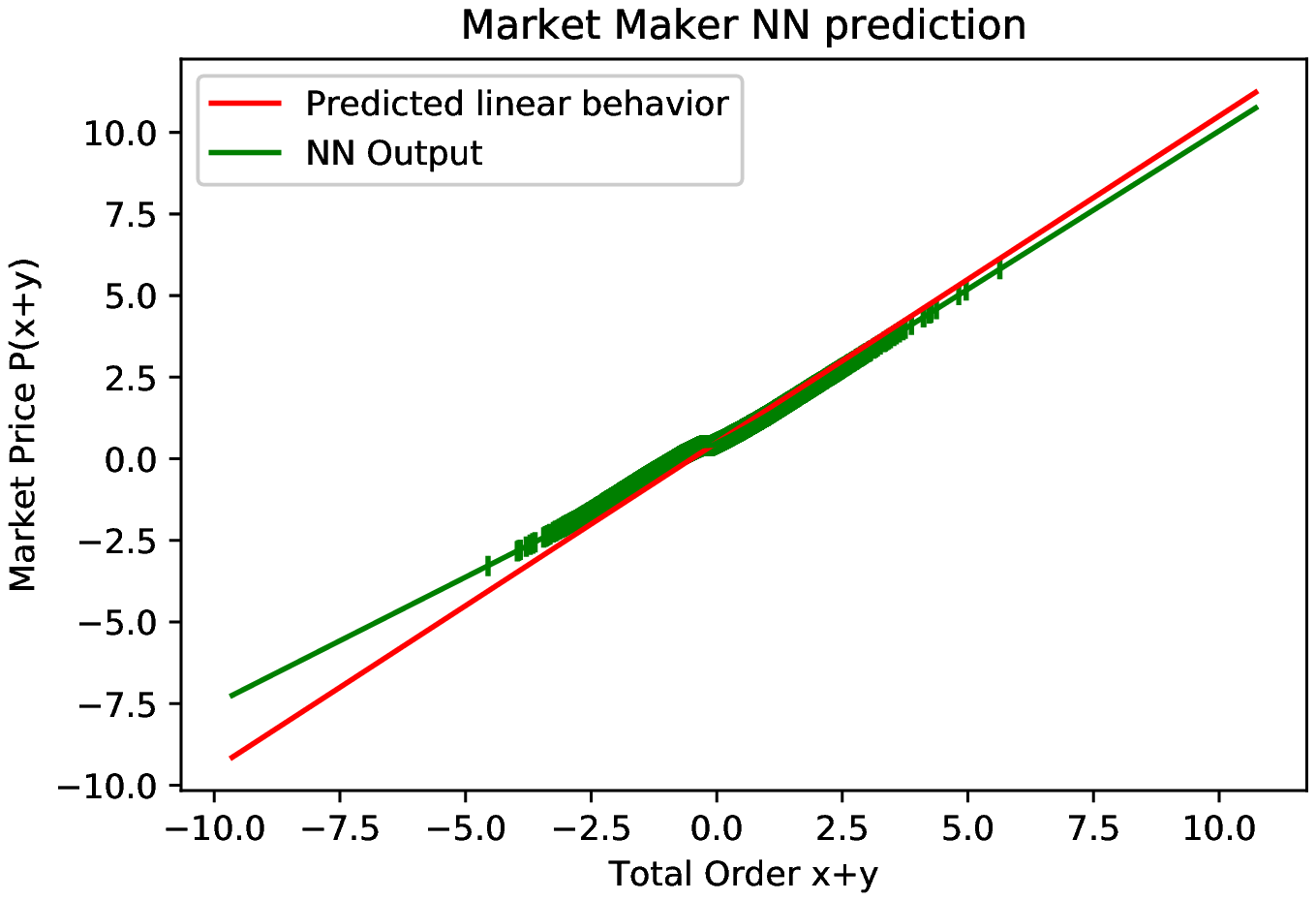}
	\label{fig:MM_15}
	\end{subfigure}%
	\begin{subfigure}[b]{0.33\textwidth}
	\centering
	\includegraphics[width=0.95\linewidth]{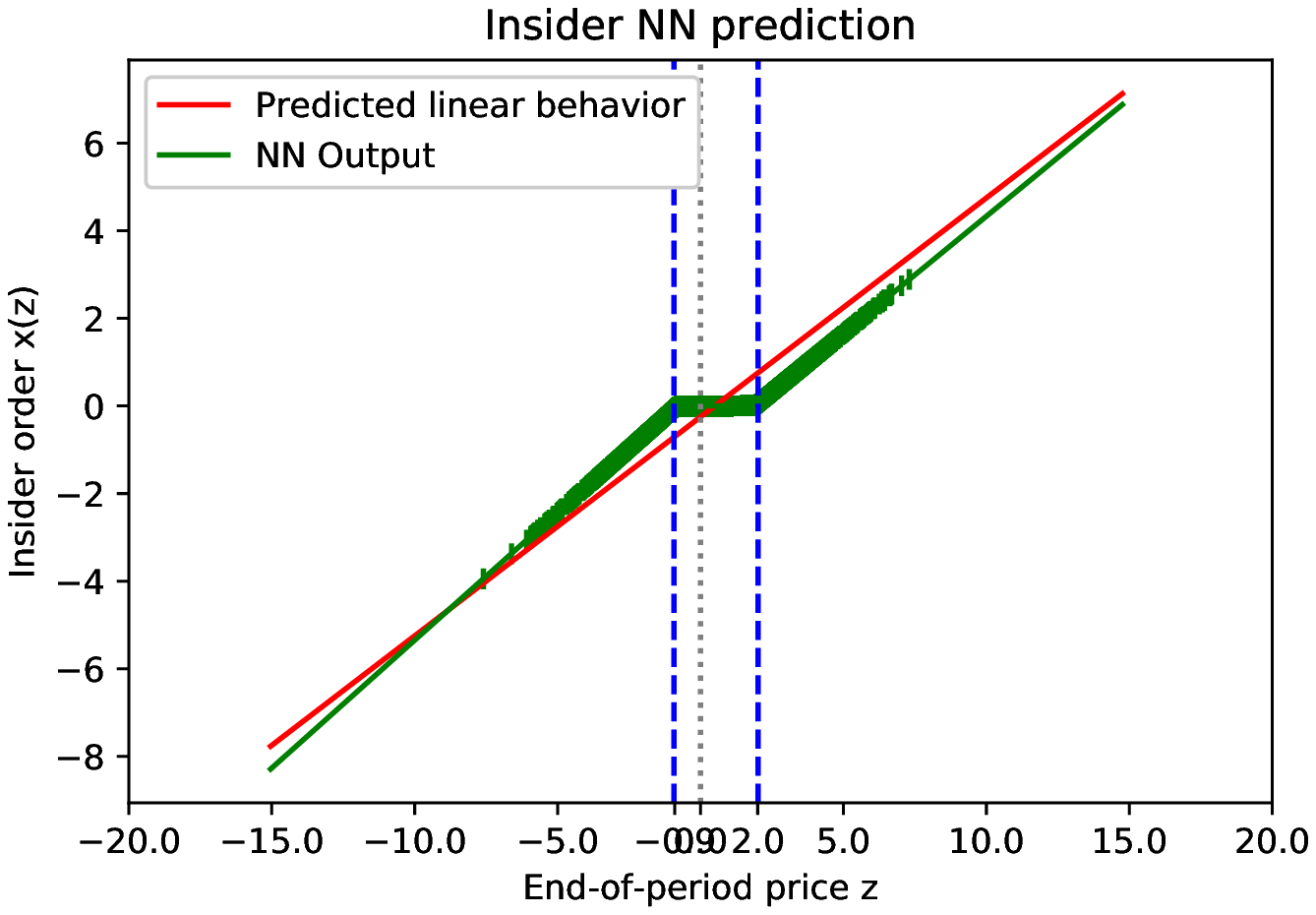}
	\label{fig:I_15_preds}
	\end{subfigure}%
	\begin{subfigure}[b]{0.33\textwidth}
	\centering
	\includegraphics[width=0.95\linewidth]{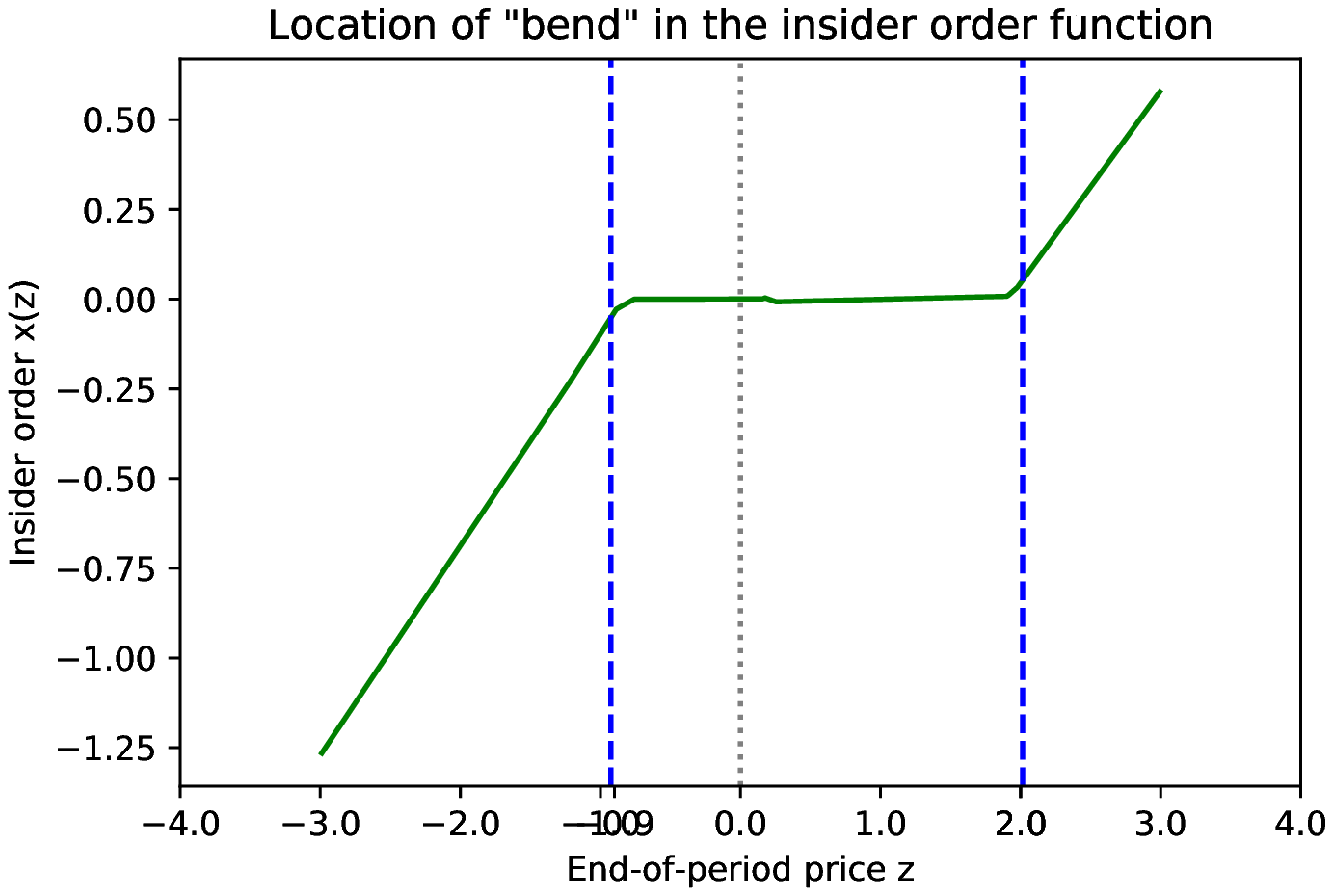}
	\label{fig:I_15_bends}
	\end{subfigure}
	
\caption{Comparing model predictions when assuming different transaction cost (epsilon) values and a fixed $\mu_z = 0.5$. Transaction cost values from top to bottom: $\epsilon = 0.1, 0.75, 1.5$.}
\label{fig:transactioncost_comparison}
\end{figure}

For our testing, we slowly increase the magnitude of the transaction cost. The results, shown in figure \ref{fig:transactioncost_comparison}, show several interesting features. The market maker pricing function stays at the linear optimum through all of our tests. For small transaction costs of $0.01$ and $0.1$, there is no visible effect on the insider's behaviour. Only when increasing the transaction cost to large values of $0.75$ or $1$ do we start to see the aforementioned `plateau' in the insider order function. The size and position of this plateau matches our hypothesis rather well, as can be seen in table \ref{table:transactioncosts} and figure \ref{fig:transactioncost_comparison}. The slight differences that are present could come from the fact that the market maker does adjust their behaviour slightly due to receiving an input order function from the insider that is not perfectly linear (i.e. that shows a plateau around zero). This would in turn influence the training and result of the insider function and could explain the small differences.

\begin{table}
\centering
\begin{tabular}{|l|l|l|l|}
\hline
$\mu_z$  & Transaction cost ($\epsilon$) & Prediction & Test result \\ \hline
0.5 & 0.01    & [0.49, 0.51]  & no visible change \\ \hline
0.5 & 0.1     & [0.4, 0.6]    & no visible change \\ \hline
0.5 & 0.5     & [0, 1]        & no visible change \\ \hline
0.5 & 0.75    & [-0.25, 1.25] & [-0.5, 1.1]       \\ \hline
0.5 & 1       & [-0.5, 1.5]   & [-0.54, 1.44]     \\ \hline
0.5 & 1.2     & [-0.7, 1.7]   & [-0.6, 1.7]       \\ \hline
1.5 & 1.2     & [0.3, 2.7]    & [0.0, 2.7]        \\ \hline
0.5 & 1.5     & [-1, 2] & [-0.9, 2.0] \\ \hline
\end{tabular}
\caption{Results of our series of tests.}
\label{table:transactioncosts}
\end{table}

\begin{figure}
\centering
	\begin{subfigure}[b]{0.5\textwidth}
	\centering
	\includegraphics[width=0.95\linewidth]{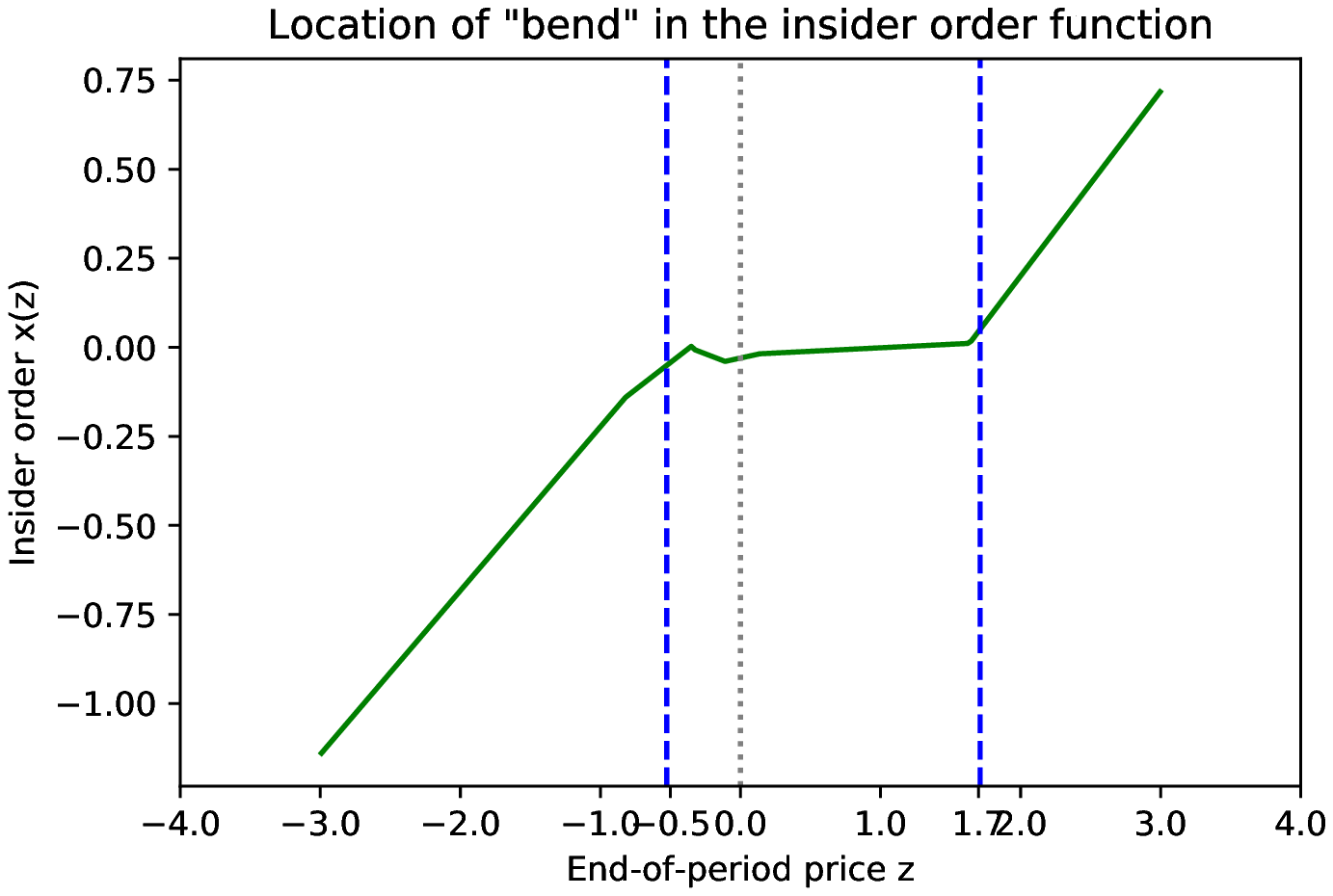}
	\end{subfigure}%
	\begin{subfigure}[b]{0.5\textwidth}
	\centering
	\includegraphics[width=0.95\linewidth]{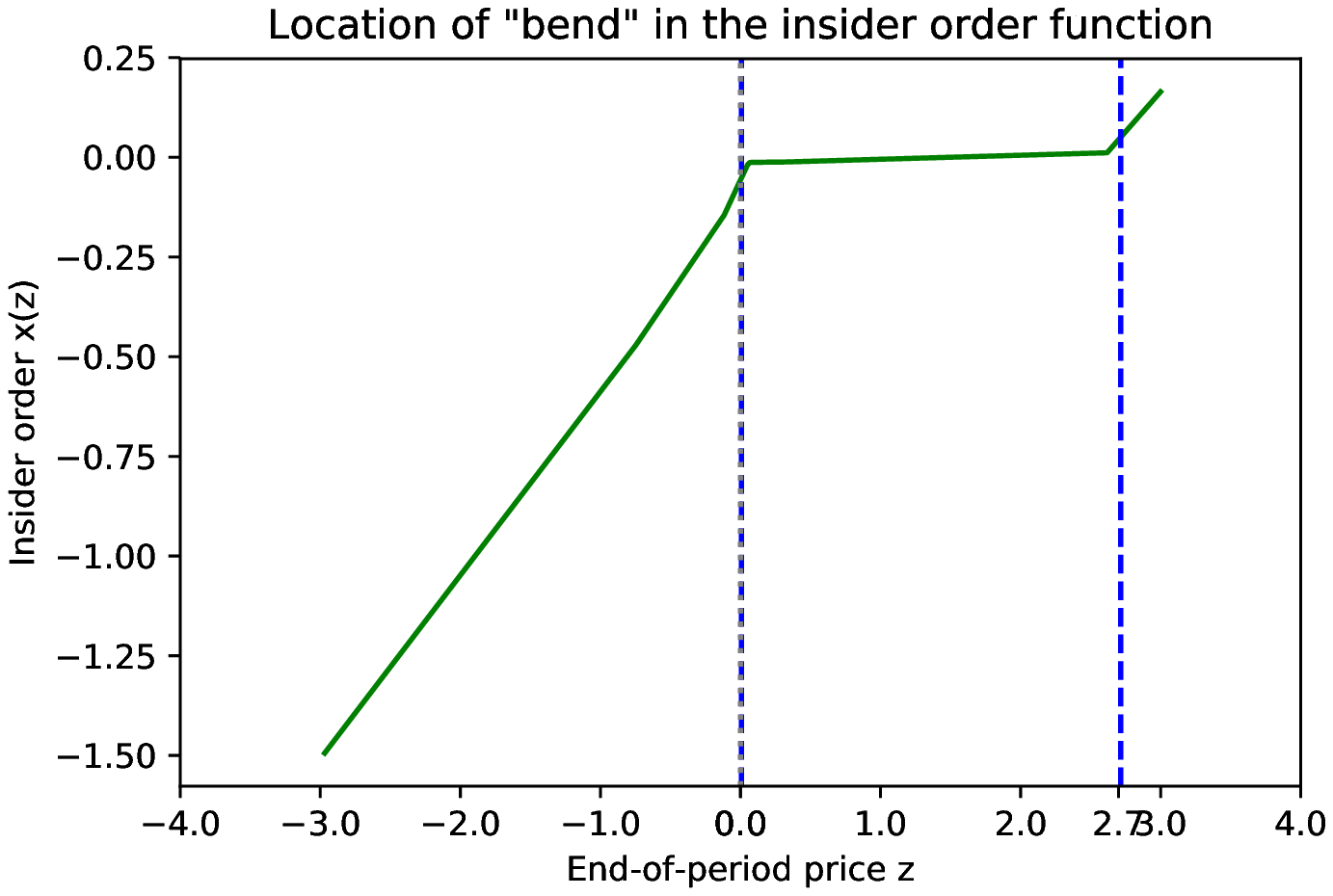}
	\end{subfigure}%
	
\caption{Left: $\mu_z = 0.5$, right: $\mu_z = 1.5$. The window where the insider function is zero remains centered around $\mu_z$, as predicted.}
\label{fig:transactioncost_shifted}
\end{figure}

In order to further confirm our hypothesis, we set $\mu_z = 1.5$. The result, shown in figure \ref{fig:transactioncost_shifted}, shows that the plateau shifts accordingly and is again roughly centered around the new value of $\mu_z$.

\section{Conclusion}

We have proven the viability of deep neural networks for finding and better understanding equilibria of the Kyle model. Our model architecture and training method leads to quick and robust convergence to Kyle's equilibrium. We can use our method to alter the model, including finding equilibria in the cases of non-normal price distributions and transaction costs. 

Our next step will be to extend our approach to the multi period Kyle model, where several auctions take place over the course of a single trading day.

\bibliographystyle{plain}
\bibliography{kylepaper_arxiv}


\end{document}